\documentclass[aps,pra,twocolumn,superscriptaddress,showpacs,amsmath,amstex,amssymb,citeautoscript]{revtex4-1}

\usepackage[T1]{fontenc}
\usepackage[utf8]{inputenc}
\usepackage{lipsum}
\usepackage{amsmath}
\usepackage{amssymb}
\usepackage{bbm}
\usepackage{braket}
\usepackage{xcolor}
\usepackage{pifont}
\usepackage{ragged2e}
\usepackage{blindtext}

\newcommand{\xmark}{\ding{55}}
\usepackage[mathscr]{euscript}
\usepackage[shortlabels]{enumitem}
\usepackage[justification=justified]{subcaption}
\usepackage{graphicx}
\usepackage{lipsum}
\allowdisplaybreaks
\usepackage{float}
\usepackage{graphicx}
\usepackage{booktabs}
\usepackage{dsfont}
\usepackage{comment}
\usepackage[normalem]{ulem}
\usepackage[colorlinks=true]{hyperref}  
\hypersetup{
    bookmarks=true,         
    unicode=false,         
    pdftoolbar=true,        
    pdfmenubar=true,        
    pdffitwindow=false,     
    pdfstartview={FitH},   
    pdftitle={Intervalley coherence and flavor polarization in three-valley moiré systems},    
    pdfauthor={Park, Classen, Scheurer},     
    pdfsubject={},   
    pdfcreator={},   
    pdfproducer={}, 
    pdfkeywords={} {} {}, 
    pdfnewwindow=true,     
    colorlinks=true,      
    linkcolor=blue, 
    citecolor=blue,        
    filecolor=magenta,     
    urlcolor=blue          
}

\newcommand{\equref}[1]{Eq.~(\ref{#1})}
\newcommand{\equsref}[2]{Eqs.~(\ref{#1}) and (\ref{#2})}
\newcommand{\secref}[1]{Sec.~\ref{#1}}
\newcommand{\figref}[1]{Fig.~\ref{#1}}
\newcommand{\refcite}[1]{Ref.~\onlinecite{#1}}

\newcommand{\tableref}[1]{Table~\ref{#1}}
\newcommand{\appref}[1]{Appendix~\ref{#1}}

\newcommand{\pdagger}{{\phantom{\dagger}}}

\renewcommand{\approx}{\simeq}

\renewcommand{\vec}[1]{\boldsymbol{#1}}

\definecolor{wrongultramarine}{rgb}{1,0.5,0}

\begin{document}

\title{Intervalley coherence and flavor polarization in three-valley moiré systems}

\affiliation{Max Planck Institute for Solid State Research, D-70569 Stuttgart, Germany}
\affiliation{School of Natural Sciences, Technische Universit\"at M\"unchen, D-85748 Garching, Germany}

\author{Jeyong Park}
\affiliation{Max Planck Institute for Solid State Research, D-70569 Stuttgart, Germany}
\affiliation{School of Natural Sciences, Technische Universit\"at M\"unchen, D-85748 Garching, Germany}
\author{Laura Classen}
\affiliation{Max Planck Institute for Solid State Research, D-70569 Stuttgart, Germany}
\affiliation{School of Natural Sciences, Technische Universit\"at M\"unchen, D-85748 Garching, Germany}
\author{Mathias S.~Scheurer}
\affiliation{Institute for Theoretical Physics III, University of Stuttgart, 70550 Stuttgart, Germany}

\begin{abstract}
We investigate interaction-induced symmetry breaking in moiré superlattices created by twisting two identical materials where the electronic low-energy degrees of freedom reside in the vicinity of the $M$ points. Based on general symmetry arguments, we identify and classify the possible candidate instabilities that, besides flavor polarized states, also involve a variety of intervalley-coherent (IVC) orders. This complexity is related primarily to the presence of three valleys, instead of the well-studied scenario of two, e.g., in graphene: IVC states can couple all three valleys identically, with a non-trivial sign structure, or even with different magnitudes. We study the energetics using an analytical strong-coupling framework and unrestricted Hartree-Fock applied to the full continuum model, with very good agreement between the two approaches. Interestingly, depending on stacking, IVC instabilities not only appear due to superexchange at moderate bandwidths, but also deep in the strong-coupling regime as a result of deviations from the flat-metric condition. Our work demonstrates that twisted $M$-point materials provide a rich playground for complex correlated physics and highlights differences and similarities to twisted multilayer graphene. 
\end{abstract}

\maketitle

\textit{Introduction.}---In recent years, twisted graphene systems have opened up a rich playground for studying correlated many-body physics \cite{Andrei2020Dec,macdonald2019bilayer}. For small relative twist angles $\theta$, their moiré band structure can be described by continuum models \cite{dos2007graphene,MeleModel,bistritzer2011moire,dos2012continuum,ContModelBalents} where the two ``valleys'' associated with the Dirac cones of the individual graphene layers become internal ``flavor’’ quantum numbers of the band structure, just like spin. Besides simply polarizing these flavors, the interaction-induced spontaneous development of intervalley coherent (IVC) order, i.e., spontaneously breaking the valley-U(1) symmetry associated with valley-charge conservation, has become one of the prime candidate instabilities of these systems \cite{PhysRevX.8.031089,PhysRevLett.124.166601,PhysRevX.10.031034,PhysRevB.103.205414,christos2022correlated,PhysRevLett.129.117602,PhysRevX.11.041063,Nuckolls2023Aug}.

Apart from creating moiré superlattices with graphene, also a large variety of other twisted moiré systems have been explored \cite{pi2026engineering,kariyado2025single,kariyado2026moir,kariyado2019flat,PhysRevB.106.075142,KoshinoTwistedTIs,PhysRevLett.122.086402,RubioGeSe,ToshiGeSe,PhysRevB.107.085127,FeSeTwisted,PhysRevB.104.035136,PhysRevB.105.165422,twistedCuprates,2024PhRvB.109h5104J,PhysRevResearch.3.033156,2024arXiv240401912W,PhysRevB.108.165422,PhysRevB.103.245206,TwistOptics,PhotonicCrystals,PhotonicCrystals2,PhysRevA.100.053604,pryds2024twisted,PhysRevB.110.125143,twistedoxides,Oxides2,Oxides3,MannhartExperiment}. These range from twisted transition metal dichalcogenides (TMDs) \cite{PhysRevLett.122.086402} and twisted surface states of topological insulators \cite{PhysRevB.106.075142,KoshinoTwistedTIs} all the way to twisting two different Bravais lattices against each other \cite{PhysRevB.110.125143,MannhartExperiment}. Most recently, twisted bilayers of materials with their low-energy degrees of freedom around their three $M$ points have received particular attention \cite{cualuguaru2025moire,lei2025moire,ingham2025moir,bao2025anisotropic}; specifically, group IV and IVB TMDs of the form 1T-MX$_2$ with $\text{M}=\text{Zr}, \text{Hf}, \text{Sn}$ and $\text{X}= \text{S}, \text{Se}$ were investigated in \cite{cualuguaru2025moire,lei2025moire}, while \refcite{ingham2025moir} focused on the effects of twisted van Hove singularities and spin-orbit coupling, e.g., in kagomé metals, and transition metal carbides were proposed in \refcite{bao2025anisotropic}. Importantly, this leads to three instead of the two valleys in the moiré band structure.

The goal of this work is to investigate the consequences for the interaction-induced instabilities arising from the increased number of valleys and their different spectral and symmetry properties in twisted $M$-point materials. 
We here use the full continuum model and study all integer fillings in contrast to a previous study at half filling \cite{de2025role} or an effective approach using a moiré-Hubbard model \cite{li2025emergent}. In particular, this will allow us to connect to and contrast with the enormous literature on twisted multi-layer graphene. We find a rich set of IVC phases which can only arise in systems with more than two valleys. We further demonstrate that---depending on stacking---an IVC phase can even dominate in the flat-band limit over competing flavor-polarized phases.

\vspace{1em}
\textit{Band structure.}---We consider the moiré superlattice formed between two identical 2D materials with hexagonal Bravais lattice and low-energy degrees of freedom located in momentum ($\vec{k}$) space in the vicinity of the three $M$ points, $M_{\eta}$, $\eta=1,2,3$. Similar to twisted bilayer graphene \cite{dos2007graphene,MeleModel,bistritzer2011moire,dos2012continuum,ContModelBalents}, the system can be approximated to be periodic for small twist angles $\theta$ between the layers. Then a continuum-model description \cite{cualuguaru2025moire,lei2025moire,ingham2025moir,bao2025anisotropic} can be employed, where the single-layer band structure around each $M$ point is written as $\epsilon_{\alpha}(\vec{k}) = \frac{(\vec{k}_{\alpha})_x^2}{2m_x} + \frac{(\vec{k}_{\alpha})_y^2}{2m_y}$; here $\vec{k}_{\alpha} = R_{\alpha}\vec{k}$, where $R_{\alpha}$ rotates 2D vectors by angle $\alpha$, and $\alpha=\frac{2\pi}{3}(\eta-1) \pm \theta/2$ with $\pm$ referring to the two layers. The presence of two different effective masses, $m_x \neq m_y$, is related to the low rotational symmetry around the $M$ points.

For small $\theta$, the emergent moiré lattice only couples  $M_\eta$ points with the same $\eta$ in the two layers. This leads to the emergence of three valleys in the low-energy band structure, to be contrasted to two valleys for $K$-point materials (like graphene). For instance, the Bloch Hamiltonian for valley $\eta=1$ and spin $s=\uparrow,\downarrow$
can be written in momentum-space as
\begin{align}
[h_{s,1}]_{\vec{k},\vec{k}'}= \left (
\begin{array}{cc} [\epsilon_{-\theta/2}(\vec{k})]\delta_{\vec{k},\vec{k}'} & T_{\vec{k},\vec{k}'}  \\
           T^{\dagger}_{\vec{k},\vec{k}'} &[\epsilon_{\theta /2}(\vec{k})]\delta_{\vec{k},\vec{k}'}
             \end{array} \right) \label{eq:Ham}.
\end{align}
Since we assume that spin-orbit coupling is negligible, which is a good approximation for candidate materials like $\textrm{SnSe}_2$ or $\textrm{ZrS}_2$, the Hamiltonian is diagonal in spin space and independent of $s$.  The Hamiltonian in the other two valleys $\eta=2,3$ is related to \equref{eq:Ham} by three-fold rotational symmetry ($C_{3z}$) along the $z$ axis. Importantly, the emergent moiré superlattice couples different momenta as described by $T_{\vec{k},\vec{k}'}=\sum^2_{j=1}(w_j\delta_{\vec{k}',\vec{k}-\vec{q}_j}+w^*_j\delta_{\vec{k}',\vec{k}+\vec{q}_j})$. Here, $\vec{q}_1 = K_{\theta}(\frac{\sqrt{3}}{2},0)$ and $\vec{q}_2 = K_{\theta}(0,\frac{3}{2})$ are the two momenta corresponding to the dominant interlayer tunneling processes, with $K_{\theta} = 8\pi \ \textrm{sin}(\theta/2)/{3a}$; $a$ is the lattice constant of the underlying materials. The tunneling amplitudes $w_j$ depend on the microscopic details of the materials considered.

\begin{figure}[t]
	\centering
	\includegraphics[width= \linewidth]
{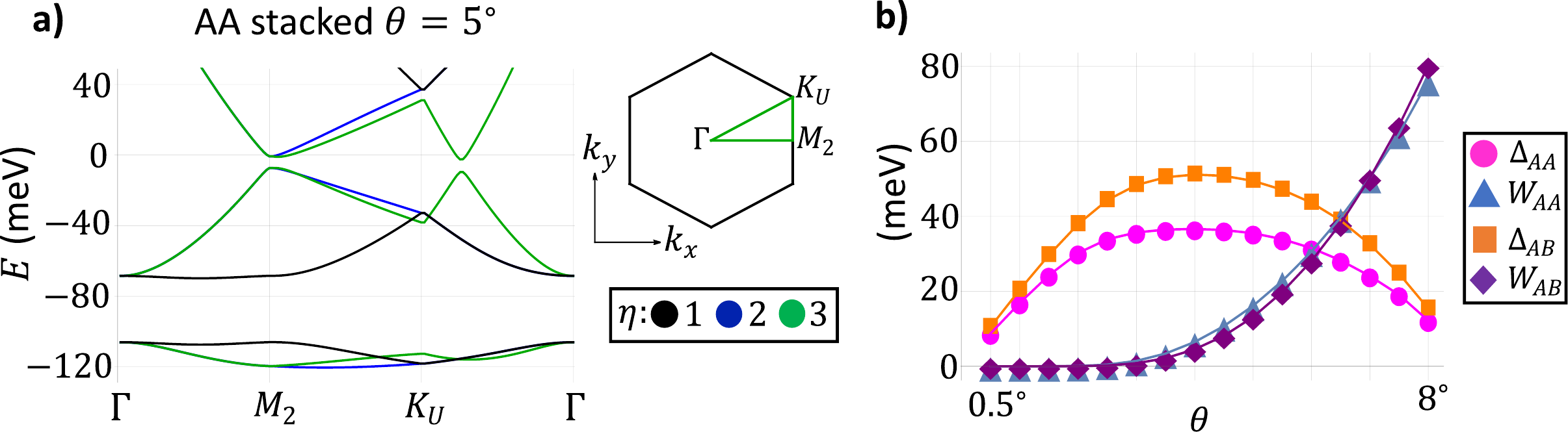}
	\caption{{\bf Single-particle physics.}
\justifying
 (\textbf{a})~Moiré band structure in all three valleys $\eta=1,2,3$ for AA-stacking and twist angle $\theta=5^{\circ}$ along the momentum path $\Gamma-M_2-K_U-\Gamma$. (\textbf{b})~Twist-angle dependence of the band gap $\Delta_{AA}, \Delta_{AB}$ between the first and second band in each valley and the bandwidth $W_{AA}, W_{AB}$ of the lowest band for both stackings. See main text for parameters and model. 	}
	\label{Fig1:SPT3V}
\end{figure}

Unless stated otherwise, we use $m_x= 0.21, \  m_y= 0.73$ in units of the bare electron mass, which corresponds to $\textrm{SnSe}_2$. Depending on the stacking structure of $\textrm{SnSe}_2$, we use  $(w_1,w_2) = (66.38 i + 88.80, -18.94)$ and $(w_1,w_2) = (-77.8, 27.04)$ in units of meV for AA and AB stacking, respectively, which have been extracted from density functional theory \cite{cualuguaru2025moire}. We show the band structure along a one-dimensional cut through the resultant hexagonal moiré Brillouin zone in \figref{Fig1:SPT3V}(a) for a twist angle of $\theta=5^{\circ}$. The plot refers to AB stacking but AA stacking leads to qualitatively similar features. In both cases, the lowest band is separated from the others by a band gap $\Delta$ that is much larger than its band width $W$ in an extended range of small $\theta$, as can be seen in \figref{Fig1:SPT3V}(b). Interestingly, $W$ decreases monotonically as $\theta$ is lowered, while $\Delta$ shows non-monotonic behavior.

\vspace{1em}
\textit{Symmetries.}---The moiré Bloch Hamiltonian introduced above is invariant under time-reversal ($\Theta$), which acts as complex conjugation within each valley. As already mentioned, it is also invariant under $C_{3z}$, corresponding to a cyclic permutation of the valleys, which together with $C_{2x}$ generates the point group $D_3$. For real $w_j$ (AB stacking), there is also an intravalley two-fold rotational symmetry, $C_{2z}$, promoting the point group to $D_6$, which contains both $C_{2x}$ and $C_{2y}$. In addition, the Bloch Hamiltonian is diagonal in the valley index such that the electronic charge is conserved separately in each valley, which can be expressed as the symmetry $[\textrm{U}(1)]_{\textrm{V}}^3 \equiv [\textrm{U}(1)]_{\textrm{V}}\times [\textrm{U}(1)]_{\textrm{V}}\times [\textrm{U}(1)]_{\textrm{V}}$. In fact, since we do not include spin-orbit coupling, the model is further invariant under a separate SU(2)$_\text{S}$ spin rotation in each valley, which we write as $[\textrm{SU}(2)]_\text{S}^3$. When studying correlations, we will focus on interactions that preserve these symmetries.

\vspace{1em}
\textit{Possible correlated orders.}---To address interaction-induced instabilities, we start with a classification purely based on symmetries. Since the main modification relative to commonly studied moir\'e systems is the presence of three instead of two valleys, we first focus entirely on the valley degree of freedom and reinstate the spin later. To capture the different symmetry-breaking channels, let use define the matrix-valued ($3 \times 3$) order parameter $\vec{\phi}_{\eta,\eta'}$ in valley space, which couples to the low-energy fermions as $\Delta H = -\sum_{\vec{k};\eta,\eta'} c^\dagger_{\vec{k},\eta} \vec{\phi}_{\eta,\eta'} c^\pdagger_{\vec{k},\eta'}$; here, $c^\dagger_{\vec{k},\eta}$ creates an electron in the band closest to the Fermi level in valley $\eta$. Note that Hermiticity requires $\vec{\phi}^\dagger = \vec{\phi}$.

The form of $\Delta H$ defines the action of the aforementioned symmetries $\Theta$, $C_{3z}$, $[\textrm{U}(1)]_{\textrm{V}}^3$, and the two-fold rotations (depending on stacking) on the order parameter $\vec\phi$. We write down the most general free energy up to quartic order in $\vec{\phi}$ consistent with these symmetries \cite{SIref} and investigate the set of all possible distinct minima. Note that all such $\vec{\phi}$ will always preserve $C_{2z}$ (which acts trivially in valley space), making the resulting set of possible instabilities identical for the two stackings, although their relative energetics does depend on the stacking order, as we will see below.
We summarize the candidate orders as follows
    \begin{align}
    \begin{split}
\textrm{VP}: \vec{\phi}_{\eta,\eta'} &= W_{\eta}\delta_{\eta,\eta'}, \label{DescriptionOfStates}
    \end{split} \\
    \begin{split}
    \textrm{IVC}^3_+: \vec{\phi}_{\eta,\eta'}&=Re^{i\varphi_{\eta\eta'}}\bar{\delta}_{\eta,\eta'}, \  \varphi_{123}=0,\nonumber
    \end{split} \\
     \begin{split}
   \textrm{IVC}^3_-: \vec{\phi}_{\eta,\eta'}&=Re^{i\varphi_{\eta\eta'}}\bar{\delta}_{\eta,\eta'}, \ \varphi_{123}=\pi\nonumber,
    \end{split} \\
    \begin{split}
      \textrm{NIVC}^1: \vec{\phi}_{\eta,\eta'}&=Y_\eta\delta_{\eta,\eta'} +R[e^{i\varphi_{12}}\delta_{\eta,1}\delta_{\eta',2}+ (1\leftrightarrow 2)], \nonumber
    \end{split} \\
    \begin{split}
      \textrm{NIVC}^3: \vec{\phi}_{\eta,\eta'}&=W_\eta\delta_{\eta,\eta'} +R_{\eta\eta'}e^{i\varphi_{\eta\eta'}}\bar{\delta}_{\eta,\eta'}, \  \varphi_{123}=0,\pi, \nonumber
    \end{split} 
   \end{align}
where $R, W_\eta,Y_\eta,R_{\eta\eta'}$ are real, positive parameters with (w.l.o.g.) $W_1=W_2>W_3$, $Y_1\geq Y_2>Y_3$ and  $R_{12}\neq R_{23}= R_{31}$. We further defined $\bar{\delta}_{\eta,\eta'}=1-\delta_{\eta,\eta'}$. In principle, the phases $\varphi_{\eta\eta'} = -\varphi_{\eta'\eta}$ can be chosen arbitrarily, except for the indicated constraint where $\varphi_{123}\equiv \varphi_{12}+\varphi_{23}+\varphi_{31} \ (\textrm{mod} \ 2\pi)$. This freedom is related to the $[\textrm{U}(1)]_{\textrm{V}}^3$ symmetry, which can be used to change two of the three independent phases in the off-diagonal components of $\vec{\phi}$, while their sum, $\varphi_{123}$, is invariant under $[\textrm{U}(1)]_{\textrm{V}}^3$.

The first state in \equref{DescriptionOfStates} corresponds to a spontaneous imbalance of the filling of the three valleys and is hence referred to as valley polarized (VP). As opposed to graphene-based or ($K$-valley) TMD-based moiré systems, such a state does not break $\Theta$ but instead breaks $C_{3z}$ symmetry. While the VP state preserves $[\textrm{U}(1)]_{\textrm{V}}^3$, the remaining options in \equref{DescriptionOfStates} do break it and are, hence, referred to as IVC states.  
We further distinguish between IVC states (IVC$^3_{\pm}$) that couple all three valleys with equal magnitude and preserve $C_{3z}$ (in the case of IVC$^3_{-}$, modulo $[\textrm{U}(1)]_{\textrm{V}}^3$) and those that do not (the ``nematic'' NIVCs). Among this latter class of IVC states, we distinguish between NIVC$^{1}$ where only one pair of valleys develops coherence, leaving one valley-charge $\textrm{U}(1)_{\textrm{V}}$ symmetry untouched (besides the global $\textrm{U}(1)_\text{C}$, associated with total charge conservation), and the remaining NIVC$^3$ where the magnitude of coherence differs between pairs of valleys. 
While the two options $\varphi_{123}=0,\pi$ for the NIVC$^3$ state appear to parallel the distinction of IVC$^3_\pm$, these two constraints lead to exactly the same symmetries for NIVC$^3$. They can thus mix and we refer to them as one phase (NIVC$^3$) in \equref{DescriptionOfStates} and in the following. 

Note that despite the constraints $\varphi_{123} = 0,\pi$, the $[\textrm{U}(1)]_{\textrm{V}}^3$ symmetry allows to make $\vec{\phi}_{\eta,\eta'}$ entirely real. This implies that the corresponding $\vec{\phi}$ is even under $\Theta$. However, it is important to keep in mind that performing a general $[\textrm{U}(1)]_{\textrm{V}}^3$ transformation will lead to another $\vec{\phi}$ that breaks time-reversal symmetry. This is in sharp contrast to $K$-valley systems where $\Theta$ and changing the relative phase  between the valleys commute. For us, it means that lattice effects, breaking $[\textrm{U}(1)]_{\textrm{V}}^3$ down to $\mathbb{Z}^3_2$, are necessary to determine the time-reversal properites of the IVC orders above. Upon their inclusion, we find \cite{SIref} that only entirely real or imaginary off-diagonal components $\vec{\phi}_{\eta,\eta'\neq \eta}$ are possible, such that each of the IVC states in \equref{DescriptionOfStates} decays into two physically distinct options---a time-reversal symmetric charge modulation on sites or bonds and a time-reversal-symmetry-breaking loop-current texture. Interestingly, these pairs of states are predicted to be asymptotically degenerate in the limit where the continuum symmetries $[\textrm{U}(1)]_{\textrm{V}}^3$ become exact. 
    
\vspace{1em}
\textit{Strong-coupling analysis.}---With the general symmetry classification of instabilities in the valleys at hand, we next address their energetics. We will consider a repulsive density-density interaction projected onto the lowest band in each valley for each spin flavor [cf.~the lowest band per spin and valley in \figref{Fig1:SPT3V}(a)]. The associated contribution to the Hamiltonian reads
\begin{equation}
    H_V = \frac{1}{2N} \sum_{\vec{q}} V_{\vec{q}} :\rho_{\vec{q}} \rho_{-\vec{q}}:, \quad V_{\vec{q}} > 0, \label{InteracingPartOfHam}
\end{equation}
where $\rho_{\vec{q}} = \sum_{\vec{k}}\sum_{\eta,s} c^\dagger_{\vec{k}+\vec{q},\eta,s} \Lambda_{\eta}(\vec{k},\vec{q}) c^\pdagger_{\vec{k},\eta,s}$ is the density operator and $:$ indicates normal ordering; here, $\Lambda_{\eta}(\vec{k},\vec{q}) = \braket{u_\eta(\vec{k}+\vec{q})|u_\eta(\vec{k})}$ are the form factors associated with the projection, where $\ket{u_{\eta}(\vec{k})}$ are the Bloch states of the active bands that follow from \equref{eq:Ham}. Our goal is to identify the energetically favored ground state for all non-trivial integer filling fractions $\nu=1,2,3,4,5$, defined via
\begin{equation}
    (P_{\vec{k}})_{s,\eta;s',\eta'} = \langle c^{\dagger}_{\vec{k},s,\eta}c^\pdagger_{\vec{k},s'\eta'}\rangle, \,\,\nu = \frac{1}{N} \sum_{\vec{k}} \text{tr}[P_{\vec{k}}]. \label{DefinitionOfP}
\end{equation}
Inspired by the success of this approach in moiré systems with two valleys \cite{PhysRevX.10.031034,PhysRevB.103.205414,PhysRevLett.122.246401,christos2022correlated,7z4z-vlj8}, we start from the strong-coupling or, equivalently, flat-band limit where the full many-body Hamiltonian is $H=H_V$. It is easy to show \cite{SIref} that any flavor-polarized product state, characterized by a diagonal $P_{\vec{k}} = P$ in the basis of \equref{DefinitionOfP} with $\nu$ entries $1$ on the diagonal and the remaining being $0$ (such that $P^2 = P$) is an exact ground state of \equref{InteracingPartOfHam} if the flat-metric condition (FMC),
\begin{equation}
    \Lambda_\eta(\vec{k},\vec{G}) = f(\vec{G}), \label{FMCMainText}
\end{equation}
applies. This type of flavor-polarized state must necessarily be VP, except at $\nu=3$ where there is also the possibility of occupying one spin flavor in each of the three valleys. This configuration, however, must necessarily be spin polarized and will thus be referred to as SP in the following. Note that, when spin is included, the labels of states (like SP) here actually refer to the equivalence classes of states generated by acting with transformations from $[\textrm{SU}(2)]_\text{S}^3$ on them. For instance, in the case of SP, this means that the spin-quantization axis in each valley can be chosen arbitrarily. 

Importantly, these flavor-polarized states do not include IVC states, which make up the majority of the candidate orders in \equref{DescriptionOfStates}. The IVC states become energetically competitive if we further assume that the form factors are close to obeying $\Lambda_{\eta}(\vec{k},\vec{q})=\Lambda(\vec{k},\vec{q})$. In this limit, $H_V$ is invariant under any global U(6) symmetry in flavor space. This means that any product state with a $\vec{k}$-independent $P_{\vec{k}} = P$ in \equref{DefinitionOfP} and obeying $P^2=P$ defines a degenerate ground state. Focusing first on $\nu=1$, it is straightforward to construct a $P$ with these properties for every order in \equref{DefinitionOfP}, except for IVC$^3_-$, which is thus shown to be incompatible with $\nu=1$ in the strong coupling limit. This is intuitively clear since the spectrum of $-\vec{\phi}$ for IVC$^3_-$ is of the form $(\epsilon_1,\epsilon_1,\epsilon_2)$ per spin with $\epsilon_2 > \epsilon_1$, which cannot lead to an insulator at $\nu=1$.

In line with the above arguments, deviations from $\Lambda_{\eta}(\vec{k},\vec{q})=\Lambda(\vec{k},\vec{q})$ favor the VP state. In turn, treating violations of the FMC in \equref{FMCMainText} in first-order perturbation theory uniquely favors the IVC$^3_+$ state over all candidate orders in \equref{DescriptionOfStates} at $\nu=1$. Since the degenerate ground states in the U(6)-symmetric limit are all product states, first order perturbation theory is equivalent to evaluating the Hartree-Fock energy functional. In this interpretation, the energetic advantage of IVC$^3_+$ over the other states comes from the Hartree term. Intuitively, the non-trivial $\eta$ dependence of $\Lambda_\eta(\vec{k},\vec{G})$ [violating \equref{FMCMainText} in addition to U(6)] cancels out in the Hartree-energy only for the IVC$^3_+$ state since its $P$ has the same value for all components, $(P)_{s,\eta;s',\eta'} = 1/3$. All other states pay an energy penalty in the Hartree energy when \equref{FMCMainText} is violated.
In addition, the IVC$^3_-$ avoids this energy penalty and is thus expected to be favored at $\nu=2$ over the spin-unpolarized form of the other candidates in \equref{DescriptionOfStates}. This includes the IVC$^3_+$ state, since IVC$^3_-$ also gains energy from the Fock term when $\Lambda_{\eta}(\vec{k},\vec{q}) \neq \Lambda(\vec{k},\vec{q})$ depends on $\eta$, breaking the U(6) symmetry. 

\begin{figure*}[t]
	\centering
	\includegraphics[width= \linewidth]{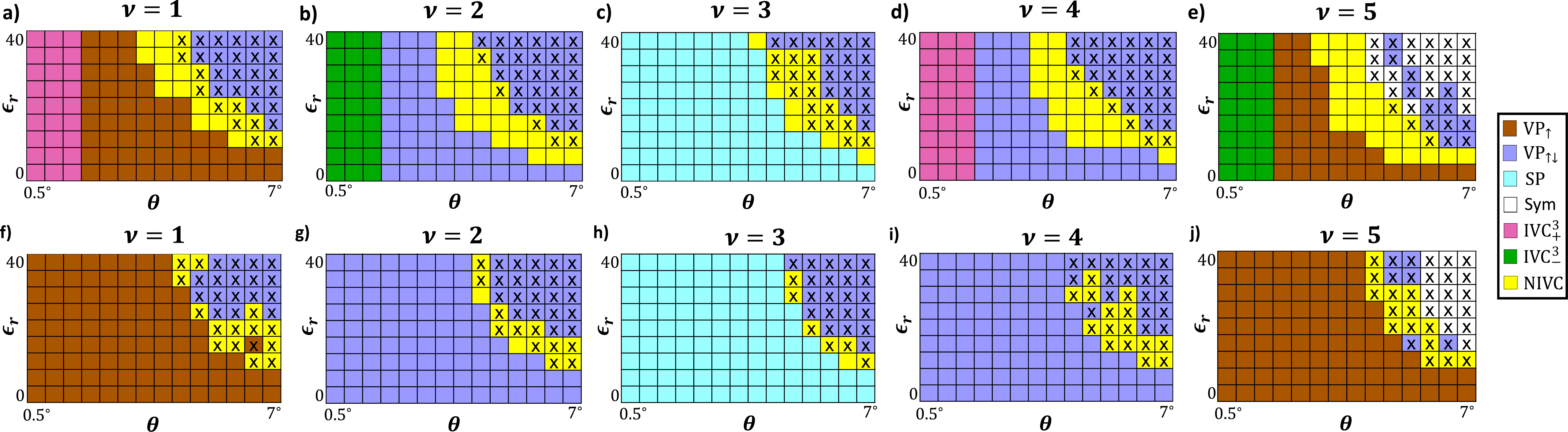}
	\caption{{\bf Hartree-Fock phase diagram.}
\justifying
In each panel, we show the character of the ground state (color, see legend) at the indicated filling $\nu$ as a function of twist angle $\theta$ and relative permittivity $\epsilon_r$, which parametrizes the interaction strength $V_{\vec{q}} \propto 1/\epsilon_r$ in \equref{InteracingPartOfHam}. The first (\textbf{a-e}) and second line (\textbf{f-j}) refers to AA and AB stacking, respectively. The ground states involve valley- (VP) and spin-polarized (SP) states, IVC phases (IVC$^3_{\pm}$ and NIVC), as well as a symmetry unbroken state (Sym). See main text for precise definition. A cross (no cross) stands for gapless (gapped) phase. The system size is $12\times12$. 
}
	\label{Fig2:HFphasediagram}
\end{figure*}

Taken together, we see that either a VP (for $\nu=3$, SP) or the IVC$^3_\pm$ ($+$ for $\nu=1$, and $-$ for $\nu=2$) state is realized in the flat-band [large $V_{\vec{q}}$ or small $\theta$, cf.~\figref{Fig1:SPT3V}(b)] limit, depending on whether the deviations from the U(6) symmetric limit of the form factors at $\vec{q}\neq \vec{G}$ or deviations from the FMC (\ref{FMCMainText}) are larger. 
As we will see below, the strong-coupling behavior for $\nu>3$ in the spinful case can be thought of as these states in one spin flavor with additional occupation of the other spin.

When also including finite bandwidth---in our current description captured by the additional contribution
\begin{equation}
    H_0 = \sum_{\vec{k}} \sum_{\eta,s} \epsilon_{\vec{k},\eta} c^\dagger_{\vec{k},\eta,s} c^\pdagger_{\vec{k},\eta,s} \label{H0Part}
\end{equation}
to the Hamiltonian---the IVC states are favored over the flavor-polarized phases. This is because they are not eigenstates of $H_0$ and can, thus, benefit from the energetic lowering in second-order perturbation theory, similar to the well-known superexchange mechanism in the Hubbard model. As such, for larger $\theta$, we expect a transition to IVC orders.

\vspace{1em}
\textit{Hartree-Fock numerics.}---We next complement these analytical considerations by a Hartree-Fock approach, which is motivated by the success of this method in graphene moiré systems \cite{kwan2025mean,PhysRevB.102.205111} and the fact that the states we obtained above are close to product states. To this end, we start from $H=H_0 + H_V$, see \equsref{InteracingPartOfHam}{H0Part}, and perform a mean-field decoupling; apart from restricting ourselves to ground states that preserve (the emergent moiré) translational symmetry, we allow for arbitrary $P_{\vec{k}}$, as defined in \equref{DefinitionOfP}, including metallic states where the $\vec{k}$-resolved filling $\text{tr}[P_{\vec{k}}]$ depends on $\vec{k}$. We refer to \cite{SIref} for further details, including the subtraction point scheme. For concreteness, we choose \equref{InteracingPartOfHam} to represent the double-gate screened Coulomb potential, such that $V_{\vec{q}} = \frac{e^2}{2\epsilon_r \epsilon_0 |\vec{q}|}\textrm{tanh}(|\vec{q}| d)$, with electron charge $e$, screening length $d$, dielectric constant $\epsilon_0$ and relative permittivity $\epsilon_r$. In the following, we fix $d=40 \ \textrm{nm}$ and tune $\epsilon_r$ to vary the strength of the interaction. In addition, to obtain the data with $\epsilon_r=0$, we set the single-particle dispersion to zero, which corresponds to the flat-band limit.

The resultant phase diagrams for all non-trivial integer fillings and for both AA and AB stackings are shown in \figref{Fig2:HFphasediagram}. Apart from a symmetry-unbroken solution, denoted by ``Sym'', the phases that appear are those listed in \equref{DescriptionOfStates} and the SP state already introduced above. With the additional spin degree of freedom, we here further distinguish between two VP states: we denote by VP$_\uparrow$ (VP$_{\uparrow\downarrow}$) the class of VP states that cannot (can) be transformed into a spin-unpolarized phase by means of the $[\textrm{SU}(2)]_\text{S}^3$ symmetry. For instance, filling one spin and one valley at $\nu=1$ constitutes a VP$_{\uparrow}$ state, while having both spin species occupied in a single valley is a VP$_{\uparrow\downarrow}$ state (so is filling two out of three valleys at $\nu=2$, independent of their spin orientation). We refer to \cite{SIref} for a detailed description of how we identify the different states and their order parameters. 
In line with our symmetry discussion above, we obtain that the NIVC$^3$ with $\varphi_{123}=0$ and $\varphi_{123}=\pi$ indeed mix in our numerics across different momenta. Numerically, we find the energies of this NIVC$^3$ and of the NIVC$^1$ state to be extremely close. Since slight changes in the model can thus tip the balance in favor of one or the other and since they both describe nematic, i.e., $C_{3z}$-breaking, IVC order, we just indicate them with a single label, NIVC, in the phase diagram. 

Overall, the Hartree-Fock results agree very well with our analytical predictions. We first note that only the VP state or IVC$^3_{\pm}$ orders appear in the strong-coupling regime (small $\theta$ and $\epsilon_r$). It is interesting that the IVC$^3_{\pm}$ orders only appear for AA stacking. As explained above, the emergence of these states is driven by the violation of the FMC, entering the Hartree term. We have checked in our numerics that, indeed, IVC$^3_{\pm}$ becomes disfavored when removing the Hartree term. Furthermore, we find by explicit evaluation of the form factors that the deviation from the FMC becomes more pronounced at very small $\theta$ where the FMC also tends to be more strongly violated for AA stacking.
We also see that, in agreement with our analysis above, IVC$^3_{+}$ wins for $\nu=1$ and IVC$^3_{-}$ for $\nu=2$. In turn, for $\nu=3$, we only obtain the flavor-polarized SP state in the strong-coupling regime. This is natural since the eigenvalues of $-\vec{\phi}$ for IVC$^3_{\pm}$ are not consistent with an insulator at $\nu=3$. For $\nu = 4$ and $\nu=5$, one can think of the strong-coupling IVC$^3_{\pm}$ orders as follows: one spin flavor has all three valleys fully occupied (like the SP at $\nu=3$) which is combined with, respectively, IVC$^3_{+}$ and IVC$^3_{-}$ in the other spin flavor. 
Upon tuning away from the strong-coupling limit (increasing $\theta$ and $\epsilon_r$), we do indeed stabilize IVC orders for both stackings and all fillings, as a result of the aforementioned superexchange mechanism. Our numerics reveal that an NIVC state, instead of IVC$^3_{\pm}$, is energetically favored. When reducing correlation effects even further, we leave the regime of validity of the perturbative approach presented above and also obtain metallic states, as indicated by the crosses in \figref{Fig2:HFphasediagram}. These metallic phases can either be Sym, NIVCs, or VP$_{\uparrow\downarrow}$ states. For additional Hartree-Fock data, including spectra, see \cite{SIref}.

\vspace{1em}
\textit{Conclusion.}---We studied spontaneous symmetry breaking in a continuum model for twisted $M$-point materials, using a combination of symmetry arguments, an analytical strong-coupling study, and unrestricted Hartree-Fock numerics. Our work reveals crucial differences compared to the more commonly-studied $K$-point materials with only two valleys such as a particularly rich set of intervalley-coherent orders. 

\textit{Note added.} In the final steps of the preparation of our manuscript, two companion papers \cite{cualuguaru2026mixed, vasiliou2026hidden} appeared on arXiv, which also deal with correlations in twisted $M$-point materials. In these works, a different technique (Monte Carlo) and theoretical model (moiré Hubbard) is used.

\textit{Acknowledgments.}--- We thank Johannes S. Hofmann for fruitful discussions. M.S.S.~further acknowledges discussions and a previous collaboration on $M$-point moiré systems with H.~Scammell and J.~Ingham. J.Y.P acknowledges Kwanjeong Educational Foundation for the support. L.C. was funded by the European Union (ERC-2023-STG, Project 101115758 - QuantEmerge). M.S.S.~also acknowledges support from the European Union (ERC-2021-STG, Project 101040651---SuperCorr). Views and opinions expressed are, however, those of the authors only and do not necessarily reflect those of the European Union or the European Research Council Executive Agency. Neither the European Union nor the granting authority can be held responsible for them.

\bibliography{draft_Refs}

\clearpage
\newpage
\onecolumngrid

\begin{appendix}
\section{Symmetry analysis}\label{AppendixSym}
In this section, we discuss the symmetries of each AA and AB stacking of the system with nonzero twist angle. 
For given band index $b$, spin index $s$, valley index $\eta$, layer index $l$, moir\'e reciprocal vector $\vec{G}$ and momentum $\vec{k}\in\textrm{mBZ}$ in the moir\'e Brillouin zone (mBZ), the wavefunction $|u_{\vec{G},b,l,s,\eta}(\vec{k})\rangle$ in plane wave basis with corresponding energy $\epsilon_{b,s,\eta}(\vec{k})$ is obtained by diagonalizing Eq.\ (\ref{eq:Ham}), i.e., solving the eigenvalue equation 
\begin{equation}
    \sum _{\vec{G}'}\sum_{l'}[H_{s,\eta}(\vec{k})]_{l,l',\vec{G},\vec{G}'}u_{\vec{G'},b,l',s,\eta}(\vec{k}) = \epsilon_{b,s,\eta}(\vec{k})u_{\vec{G},b,l, s,\eta}(\vec{k}).
\end{equation}
We first discuss again the symmetries in our model and then specify their representation. The relevant generators of the symmetry group are (i) $C_{3z}$: intervalley $120^{\circ}$ rotation symmetry about the $z$ axis, (ii) $C_{2z}$: intravalley $180^{\circ}$ rotation symmetry about the $z$ axis, (iii) $C_{2x}$: intervalley $180^{\circ}$ rotation symmetry about the $x$ axis, (iv) $\Theta$: time reversal symmetry, (v) $[\textrm{U}(1)]^3_{\textrm{V}}$: inpdenendent phase rotation symmetry in each valley. For each symmetry transformation $S$, the following relation holds  
\begin{equation}
    u_{S_k\vec{G},b,l,s,\eta}(S_k\vec{k}) = \sum_{l',s',\eta'}[U_{S}]_{l,l',s,s',\eta,\eta'} u_{\vec{G},b,l',s',\eta'}(\vec{k}), \label{SymRep}
\end{equation}
with spinor transformation matrix $U_{S}$ written in layer, valley and spin space as 
\begin{subequations}
\begin{align}
\begin{split}U_{C_{3z}} = l_0 \otimes e^{i\pi s_z/3} \otimes \left (
\begin{array}{ccc} 0 & 0 & 1  \\
          1 &0 &  0 \\
          0 &  1& 0
             \end{array} \right), \ \ U_{C_{2z}} = l_0  \otimes e^{i\pi s_z/2} \otimes \mathbbm{I}_{3\times3}, \ \  U_{C_{2x}} =  l_0 \otimes e^{i\pi s_x/2} \otimes \left (
\begin{array}{ccc} 1 & 0 & 0  \\
          0 &0 &  1 \\
          0&  1& 0
             \end{array} \right)
             \end{split} \nonumber \\
             \begin{split}
              U_{\Theta}  = l_0  \otimes i s_y \mathcal{K} \otimes\mathbbm{I}_{3\times3}, \ \ U_{[\textrm{U}(1)]^3_{\textrm{V}}}  =  l_0 \otimes s_0 \otimes \left (
\begin{array}{ccc} e^{i\varphi_1} & 0 & 0  \\
          0&e^{i\varphi_2} &  0 \\
          0&  0& e^{i\varphi_3}
             \end{array} \right),
            \nonumber \end{split}
\end{align}
\end{subequations}
where we define $l_\alpha$ and $s_{\alpha} \ (\alpha \in \{0,x,y,z\})$ as Pauli matrices in layer and spin basis, respectively, $\mathcal{K}$ as the complex conjugation operator and real phase factors $\varphi_1, \varphi_2,\varphi_3$. Note that the representation $S_k$ of $S$ on momentum in \equref{SymRep} for some symmetries involves an additional shift, due to their non-symmorphic nature \cite{cualuguaru2025moire}.

We here state the spinful (relativistic) versions of those transformations, where spin and orbital degrees of freedom rotate together. Since our model has no spin-orbit coupling, the spinless variants, which are trivial in spin space, are also symmetries of the model.
Here, we choose the index of valleys such that under the $C_{2x}$ transformation for AA (AB) stacking, the intervalley transformation acts as $3\rightarrow 3$ and $1\rightarrow2$, $2\rightarrow 1$. 

Importantly, $C_{2z}$ is only preserved by the model if the tunneling parameters $w_j$ are real. This is the case for AB but not for AA stacking. Consequently, apart from moiré translations and the $[\textrm{SU}(2)]_\text{S}^3$ spin symmetry mentioned in the main text, the symmetry group for AA stacking is generated by $\{C_{3z}, C_{2x}, \Theta,[\textrm{U(1)}]^3_{\textrm{V}} \}$ and that for AB stacking by $\{C_{3z}, C_{2z}, C_{2x}, \Theta,[\textrm{U(1)}]^3_{\textrm{V}} \}$. Note that the latter contains both $C_{2x}$ and $C_{2y}$ rotational symmetries.

\section{Ginzburg-Landau analysis} \label{GLanal}

\subsection{Symmetries and Ginzburg-Landau free energy}\label{GLorders}
In this section, we detail our Ginzburg-Landau procedure. For given valley index $\eta =1,2,3$, we first define the fermion creation operator $c^{\dagger}_{\eta}$ and the order parameter $O = \sum_{\eta,\eta'}c^{\dagger}_{\eta}\vec{\phi}_{\eta,\eta'}c_{\eta'}$, with $ \vec{\phi}_{\eta,\eta'}$ in valley space. Note that since $O$ is Hermitian, $\vec{\phi}_{\eta,\eta'} =\vec{\phi}_{\eta',\eta}^*$. Our goal is to write the symmetry-allowed Ginzburg-Landau free energy of with respect to $\vec{\phi}_{\eta,\eta'}$. As discussed in Appendix \ref{AppendixSym} in detail, the symmetries present in the systems are either $\{C_{3z}, C_{2z}, C_{2x}, \Theta,U_{[\textrm{U}(1)]_{\textrm{V}}^3} \}$ or $\{C_{3z}, C_{2x}, \Theta,U_{[\textrm{U}(1)]_{\textrm{V}}^3} \}$, which depends on the stacking. However, these lead to the same free energy because the form of $\vec \phi$ makes it automatically invariant under $C_{2z}$. For the different symmetry operations, $\vec{\phi}_{\eta,\eta'}$ transforms as
\begin{enumerate}
    \item  $C_{3z}$: $\vec{\phi}_{\eta,\eta'} \rightarrow \vec{\phi}_{\eta+1,\eta'+1} \ \ [(\eta=4)\equiv(\eta=1)]$
    \item  $C_{2z}$: $\vec{\phi}_{\eta,\eta'} \rightarrow \vec{\phi}_{\eta,\eta'}$
    \item  $C_{2x}$: $\vec{\phi} \rightarrow U^{\dagger} \vec{\phi}U, \ U_{\eta,\eta'} = \delta_{\eta,1}\delta_{\eta',2}+\delta_{\eta,2}\delta_{\eta',1}+\delta_{\eta,3}\delta_{\eta',3}$
    \item $\Theta$: $ \vec{\phi}_{\eta,\eta'} \rightarrow \vec{\phi}^*_{\eta,\eta'}$
    \item $[\textrm{U}(1)]_{\textrm{V}}^3$ : $ \vec{\phi}_{\eta,\eta'} \rightarrow e^{i(\varphi_{\eta} - \varphi_{\eta'})}\vec{\phi}_{\eta,\eta'}$,
\end{enumerate}
where $\varphi_{\eta}$ is the arbitrary phase factor from the  $[\textrm{U(1)}]_{\textrm{V}}$ valley rotation. Using these symmetry transformations, we write the general symmetry-allowed Ginzburg-Landau free energy $F(\vec{\phi}_{\eta,\eta'}) \ (\eta,\eta'=1,2,3)$ of a system up to the fourth order of $\vec{\phi}_{\eta,\eta'}$. Based on the symmetry analysis for quadratic, cubic, and quartic combinations of $\vec \phi_{\eta,\eta'}$, we find 18 possible terms for the Ginzburg-Landau free energy, which we write as $F=\sum_{l=1}^{18} A_l f_l (\vec{\phi})$ with the nine component vector $\vec{\phi}\equiv(\vec{\phi}_{1,1}, \vec{\phi}_{1,2}, \vec{\phi}_{1,3}, \vec{\phi}_{2,1},\vec{\phi}_{2,2},\vec{\phi}_{2,3},\vec{\phi}_{3,1},\vec{\phi}_{3,2},\vec{\phi}_{3,3})$ and
\begin{subequations}
    \begin{align}
    \begin{split}
f_1(\vec{\phi})=\vec{\phi}^2_{1,1}+\vec{\phi}^2_{2,2}+\vec{\phi}^2_{3,3}\nonumber
    \end{split} \\
    \begin{split}
f_2(\vec{\phi})=|\vec{\phi}_{1,2}|^2+|\vec{\phi}_{2,3}|^2+|\vec{\phi}_{3,1}|^2\nonumber\nonumber
    \end{split} \\
     \begin{split}
  f_3(\vec{\phi})=\vec{\phi}_{1,1}\vec{\phi}_{2,2}+\vec{\phi}_{2,2}\vec{\phi}_{3,3}+\vec{\phi}_{3,3}\vec{\phi}_{1,1}\nonumber
    \end{split} \\
    \begin{split}
    f_4(\vec{\phi})=\vec{\phi}_{1,1}\vec{\phi}_{2,2}\vec{\phi}_{3,3} \nonumber
    \end{split} \\
    \begin{split}
    f_5(\vec{\phi})=\vec{\phi}_{1,1}(\vec{\phi}^2_{2,2}+\vec{\phi}^2_{3,3})+\vec{\phi}_{2,2}(\vec{\phi}^2_{3,3}+\vec{\phi}^2_{1,1})+\vec{\phi}_{3,3}(\vec{\phi}^2_{1,1}+\vec{\phi}^2_{2,2}) \nonumber
    \end{split} \\
    \begin{split}
    f_6(\vec{\phi})= \vec{\phi}^3_{1,1} + \vec{\phi}^3_{2,2}+\vec{\phi}^3_{3,3}  \nonumber
    \end{split}\\
    \begin{split}
    f_7(\vec{\phi})= \vec{\phi}_{1,1}(|\vec{\phi}_{1,2}|^2+|\vec{\phi}_{1,3}|^2) + \vec{\phi}_{2,2}(|\vec{\phi}_{2,3}|^2+|\vec{\phi}_{2,1}|^2)+\vec{\phi}_{3,3}(|\vec{\phi}_{3,1}|^2+|\vec{\phi}_{3,2}|^2)  \nonumber
    \end{split}\\
    \begin{split}
    f_8(\vec{\phi})= \vec{\phi}_{1,1}|\vec{\phi}_{2,3}|^2 + \vec{\phi}_{2,2}|\vec{\phi}_{3,1}|^2+\vec{\phi}_{3,3}|\vec{\phi}_{1,2}|^2 \nonumber
    \end{split} \\
    \begin{split}
    f_{9}(\vec{\phi})= \textrm{Re}[\vec{\phi}_{1,2}\vec{\phi}_{2,3}\vec{\phi}_{3,1}] \nonumber
    \end{split} \\
    \begin{split}
    f_{10}(\vec{\phi})= \vec{\phi}^4_{1,1}+\vec{\phi}^4_{2,2}+\vec{\phi}^4_{3,3}\nonumber
    \end{split}\\
    \begin{split}
    f_{11}(\vec{\phi})= \vec{\phi}^3_{1,1}(\vec{\phi}_{2,2}+\vec{\phi}_{3,3})+\vec{\phi}^3_{2,2}(\vec{\phi}_{3,3}+\vec{\phi}_{1,1})+\vec{\phi}^3_{3,3}(\vec{\phi}_{1,1}+\vec{\phi}_{2,2})\nonumber
    \end{split} \\
    \begin{split}
    f_{12}(\vec{\phi})= \vec{\phi}^2_{1,1}\vec{\phi}^2_{2,2}+\vec{\phi}^2_{2,2}\vec{\phi}^2_{3,3}+\vec{\phi}^2_{3,3}\vec{\phi}^2_{1,1}\nonumber
    \end{split} \\
    \begin{split}
    f_{13}(\vec{\phi})= \vec{\phi}^2_{1,1}\vec{\phi}_{2,2}\vec{\phi}_{3,3}+\vec{\phi}^2_{2,2}\vec{\phi}_{3,3}\vec{\phi}_{1,1}+\vec{\phi}^2_{3,3}\vec{\phi}_{1,1}\vec{\phi}_{2,2}\nonumber
    \end{split}\\
    \begin{split}
    f_{14}(\vec{\phi})= |\vec{\phi}_{1,2}|^4+|\vec{\phi}_{2,3}|^4+|\vec{\phi}_{3,1}|^4\nonumber
    \end{split} \\
    \begin{split}
    f_{15}(\vec{\phi})= |\vec{\phi}_{1,2}|^2|\vec{\phi}_{2,3}|^2+|\vec{\phi}_{2,3}|^2|\vec{\phi}_{3,1}|^2+|\vec{\phi}_{3,1}|^2|\vec{\phi}_{1,2}|^2\nonumber
    \end{split} \\
    \begin{split}
    f_{16}(\vec{\phi})= \textrm{Re}[\vec{\phi}_{1,2}\vec{\phi}_{2,3}\vec{\phi}_{3,1}] (\vec{\phi}_{1,1}+\vec{\phi}_{2,2}+\vec{\phi}_{3,3})\nonumber
    \end{split}\\
    \begin{split}
    f_{17}(\vec{\phi})= \vec{\phi}^2_{1,1}(|\vec{\phi}_{1,2}|^2+|\vec{\phi}_{3,1}|^2) + \vec{\phi}^2_{2,2}(|\vec{\phi}_{2,3}|^2+|\vec{\phi}_{1,2}|^2)+\vec{\phi}^2_{3,3}(|\vec{\phi}_{2,3}|^2+|\vec{\phi}_{3,1}|^2)\nonumber
    \end{split}\\
        \begin{split}
    f_{18}(\vec{\phi})= \vec{\phi}^2_{1,1}|\vec{\phi}_{2,3}|^2 + \vec{\phi}^2_{2,2}|\vec{\phi}_{3,1}|^2+\vec{\phi}^2_{3,3}|\vec{\phi}_{1,2}|^2\nonumber
    \end{split}
    \end{align}
\end{subequations}

\begin{figure*}[t]
	\centering
	\includegraphics[width= 1\columnwidth]{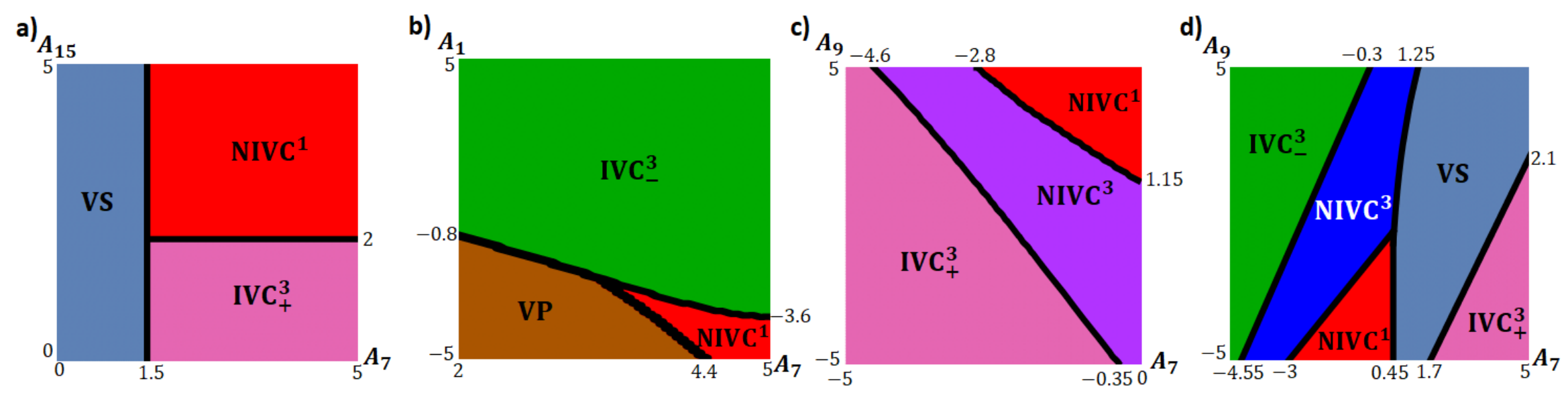}
	\caption{{\bf Ginzburg-Landau phase diagram}
\justifying
for different combinations of parameters specified in the text in Appendix \ref{GLanal}. VS stands for ``valley symmetric'', which means that no symmetry is broken; meanwhile, all other phases are defined in the main text and \tableref{SymmetriesOfStates}.
We point out that for the NIVC$^3$ state in (\textbf{c}) and (\textbf{d}), we get $\varphi_{123}=0$ and $\varphi_{123}=\pi$, respectively, and we used different colors in the phase diagram to display them.
	}
	\label{Fig2:GLphase}
\end{figure*}

\subsection{Ginzburg-Landau phase diagram}
Now, we numerically compute the $\vec{\phi}$ that minimizes $F$. To simplify our analysis, we only tune four parameters $A_1, A_{7}, A_{9}, A_{15}$ and fix the remaining parameters, because this already allows to demonstrate the appearance of all possible correlated states. We checked by varying the parameters that no additional phases appear as stable minima of the free energy $F=\sum_{l=1}^{18} A_l f_l (\vec{\phi})$.

Figure \ref{Fig2:GLphase} shows the resulting Ginzburg-Landau phase diagram for different parameter sets of $(A_1, A_7, A_{9}, A_{15})$. In Fig.\ \ref{Fig2:GLphase}(a), we show the phase diagram as a function of $(A_7,A_{15})$  fixing the remaining parameters to one. In Fig.~\ref{Fig2:GLphase}(b), we show the phase diagram as a function of $(A_1,A_{7})$ again, fixing $A_9=4$ and the remaining parameters to one, and finally in Fig.~\ref{Fig2:GLphase}(c) and (d), we vary $(A_7,A_{9})$ and either $A_{16}=-1$ or $A_{16}=1$, and all other parameters set to
\begin{subequations}
    \begin{align}
    \begin{split}
(A_{10},A_{12},A_{13},A_{14},A_{15},A_{17})=(1, 1, 1, 1, 1, 1)\nonumber
    \end{split} \\
    \begin{split}
(A_1,A_{2},A_{3},A_{4},A_{5},A_{6},A_{8},A_{11},A_{18})=(-1, -1, -1, -1, -1, -1, -1, -1, -1)\nonumber.
    \end{split}
    \end{align}
\end{subequations}

Our Ginzburg-Landau phase diagrams show that depending on the parameters (which may differ for AA and AB stacking), one of VP, $\textrm{IVC}^3_{\pm}$, $\textrm{NIVC}^1$ and $\textrm{NIVC}^3$ states can become a ground state. In Fig.\ \ref{Fig2:GLphase}(c) and (d), for NIVC$^3$ state, we used the different colors depending on whether the phase factor $\varphi_{123}=0$ (purple) or $\varphi_{123}=\pi$ (blue) (even though they cannot be distinguished by symmetry, see main text). All phase boundaries correspond to  first order phase transitions. Note that the parameters $A_1,A_2$ and $A_3$ control the symmetry breaking of VP and IVC orders on the quadratic level. For example, we can see that their relation tunes between VP and IVC$_-^3$ in Fig.~\ref{Fig2:GLphase}(b). If quadratic terms indicate both VP and IVC order, quartic terms will decide on the ground state. See Eq.~\eqref{DescriptionOfStates} in the main text for the detailed definition of each state.
The symmetries of the different candidate orders from \equref{DescriptionOfStates} of the main text and derived here, can be found in \tableref{SymmetriesOfStates}. 

\begin{table}[h!]
\centering
\caption{ {\bf{Symmetry analysis of possible Ginzburg-Landau ground states.}} \justifying $\checkmark^*$ stands for the case where the state can be made to respect the corresponding symmetry by properly choosing the rotation axis and/or the valley degrees of freedom. For all IVCs, we assume that $\vec{\phi}_{\eta,\eta'}$ has been made real upon taking advantage of the [U(1)]$_\text{V}^3$ degree of freedom. Within these conventions, some of the IVC states either explicitly preserve the symmetries ($\checkmark$ in columns 2-5), preserve it only upon combining it with a proper [U(1)]$_\text{V}^3$ transformation ($\checkmark$ in columns 6 and 7), or break it completely (\xmark~in columns 6 and 7).
In the last column, we indicate which subgroup the [U(1)]$_\text{V}^3$ valley-charge conservation symmetry is reduced to. Here, U(1)$_\textrm{C}$ and U(1)$_\textrm{V}$ corresponds to the conservation of total charge and the charge in one of the three valleys, respectively.}
\renewcommand{\arraystretch}{1.2}
\begin{tabular}{ccccc|cc|c}
\toprule
 & $C_{3z}$ & $C_{2x}$ & $C_{2z}$ & $\Theta$ & $\textrm{[U(1)]}^3_{\textrm{V}} C_{3z}$ & $\textrm{[U(1)]}^3_{\textrm{V}} C_{2x}$  & [U(1)]$_{\textrm{V}}^3$ \\
\midrule
VP     & \xmark & $\checkmark^*$  & $\checkmark$ & $\checkmark$ & \xmark & $\checkmark^*$ &  [U(1)]$^3_\text{V}$ \\
$\textrm{IVC}^3_+$      & $\checkmark$ & $\checkmark$  & $\checkmark$ & $\checkmark$ & $\checkmark$ & $\checkmark$&  U(1)$_\text{C}$ \\
$\textrm{IVC}^3_-$  & \xmark & $\checkmark$  & $\checkmark$ & $\checkmark$ & $\checkmark$ & $\checkmark$&  U(1)$_\text{C}$ \\
$\textrm{NIVC}^1$  & \xmark & $\checkmark^*$ & $\checkmark$ & $\checkmark$ & \xmark & $\checkmark^*$ & U(1)$_\text{C} \times$ U(1)$_\text{V}$ \\
$\textrm{NIVC}^3$ & \xmark & $\checkmark^*$ & $\checkmark$ & $\checkmark$ & \xmark& $\checkmark^*$  & U(1)$_\text{C}$ \\
\bottomrule
\end{tabular}
\label{SymmetriesOfStates}
\end{table}

\subsection{Lattice effects}
Importantly, for $M$-point materials, changing the relative phase between the valleys does not commute with time-reversal $\Theta$. This comes from the fact that $\Theta$ does not map between the valleys, as opposed to $K$-point materials. For the latter, $\Theta_K= \tau_x \mathcal{K}$, where $\tau_j$ are Pauli matrices in the now two valleys, such that $[\Theta_K,e^{i\varphi \tau_z}]=0$. This means that we need to take into account effects associated with the microscopic lattices of the underlying materials that make up the moiré bilayer system. Since the $M$ points are each half of a reciprocal lattice vector, translating the system by a (microscopic, non-moiré) lattice constant along one of the three elementary directions $\vec{a}_j$, $j=1,2,3$, of the underlying triangular lattice is associated with
\begin{equation}
    c_\eta \quad \longrightarrow \quad D^{(j)}_{\eta,\eta'} c_{\eta'}, \quad D^{(j)} = -\text{diag}[(-1)^{\delta_{j,1}},(-1)^{\delta_{j,2}},(-1)^{\delta_{j,3}}].
\end{equation}
So we see that the $[\textrm{U}(1)]_{\textrm{V}}^3$ symmetry is broken down to $\mathbb{Z}_2^3$ by the lattice. Focusing on the lowest-order (quadratic) additional term in the free energy, we describe the lattice effects by
\begin{equation}
    \Delta F = \delta \, \text{Re}\left[ \vec{\phi}_{1,2}^2 + \vec{\phi}_{2,3}^2 + \vec{\phi}_{3,1}^2  \right]. 
\end{equation}
Depending on the sign of the real parameter $\delta$, we thus either find $\vec{\phi}_{\eta,\eta'\neq \eta} \in \mathbb{R}$ or $\vec{\phi}_{\eta,\eta'\neq \eta} \in i \mathbb{R}$, as stated in the main text.

\section{Strong-coupling analysis}
For our strong-coupling analysis, we start by studying a density-density interaction term projected onto the single active band at the Fermi level for each valley $\eta=1,2,3$ and each spin $s=\uparrow,\downarrow$. Although the Hamiltonian was already defined in the main text, we here reiterate the key definitions for convenience of the reader. The strong-coupling Hamiltonian reads as
\begin{equation}
    H_V = \frac{1}{2N} \sum_{\vec{q}} V_{\vec{q}} :\rho_{\vec{q}} \rho_{-\vec{q}}:, \quad \rho_{\vec{q}} = \sum_{\vec{k}}\sum_{\eta,s} c^\dagger_{\vec{k}+\vec{q},\eta,s} \Lambda_{\eta}(\vec{k},\vec{q}) c^\pdagger_{\vec{k},\eta,s}, \label{HVInteraction}
\end{equation}
where $c^\dagger_{\vec{k},\eta,s}$ are creation operators of electrons in valley $\eta$ and of spin $s$ and $N$ is the system size. For now, the precise form of $V_{\vec{q}}$ is irrelevant; we will only assume that $V_{\vec{q}} >0$ and $V_{\vec{q}} = V_{-\vec{q}}$. Furthermore, $\Lambda_{\eta}(\vec{k},\vec{q})$ are the form factors defined in terms of the Bloch states $\ket{u_\eta(\vec{k})}$ of the single (spin-degenerate) active band in each valley as
\begin{equation}
    \Lambda_{\eta}(\vec{k},\vec{q}) = \braket{u_\eta(\vec{k}+\vec{q})|u_\eta(\vec{k})}.
\end{equation}
For later reference, we note that Hermiticity implies 
\begin{equation}
    \Lambda_{\eta}(\vec{k}+\vec{q},-\vec{q}) = \Lambda^*_{\eta}(\vec{k},\vec{q}) \quad \text{such that} \quad \Lambda_{\eta}(\vec{k},-\vec{G}) = \Lambda^*_{\eta}(\vec{k},\vec{G}). \label{HermiticityConstraint}
\end{equation}
Note that for AB stacking, the presence of $C_{2z}\Theta$ implies the existence of a gauge with real-valued $\Lambda_{\eta}(\vec{k,q})$ for all momenta $\vec{k}$ and $\vec{q}$, whereas for AA stacking, the form factor is in general a complex number due to the absence of $C_{2z}\Theta$ symmetry. 

In \equref{HVInteraction}, we have already used the fact that the non-interacting part of the Hamiltonian is diagonal in the valley index and does not depend on spin (since we neglect spin-orbit coupling). This alone is already sufficient to see that any flavor-polarized state is an eigenstate of $H_V$. To this end, let $\mathcal{S}$ be a set of $\nu\in\{1,2,3,4,5,6\}$ distinct combinations of the quantum numbers $(\eta,s)$ and define
\begin{equation}
    \ket{\text{FP}(\mathcal{S})} := \prod_{(s,\eta)\in \mathcal{S}} \prod_{\vec{k}} c^\dagger_{\vec{k},\eta,s} \ket{0}, \label{FPStates}
\end{equation}
where $\ket{0}$ is the empty vacuum state. This defines a state at filling fraction $\nu = |\mathcal{S}|$. Since each flavor is now either entirely empty or fully filled it holds
\begin{align}
    \rho_{\vec{q}}\ket{\text{FP}(\mathcal{S})} &= \sum_{\vec{k}}\sum_{\eta,s} c^\dagger_{\vec{k}+\vec{q},\eta,s} \Lambda_{\eta}(\vec{k},\vec{q}) c^\pdagger_{\vec{k},\eta,s}\ket{\text{FP}(\mathcal{S})} \\
    & = \delta_{\vec{q} \in \text{RL}} \sum_{\vec{k}} \sum_{(\eta,s) \in\mathcal{S}} c^\dagger_{\vec{k},\eta,s} \Lambda_{\eta}(\vec{k},\vec{q}) c^\pdagger_{\vec{k},\eta,s}\ket{\text{FP}(\mathcal{S})}, \qquad \delta_{\vec{q} \in \text{RL}} := \sum_{\vec{G}\in\text{RL}} \delta_{\vec{q},\vec{G}}, \\
    & =  \delta_{\vec{q} \in \text{RL}} \lambda(\vec{q}) \ket{\text{FP}(\mathcal{S})}, \qquad \lambda(\vec{q}) = \sum_{\vec{k}} \sum_{(\eta,s) \in\mathcal{S}}  \Lambda_{\eta}(\vec{k},\vec{q}).
\end{align}
Being an eigenstate of $\rho_{\vec{q}}$, $\ket{\text{FP}(\mathcal{S})}$ is also an eigenstate of $H_V$. However, this does not mean that it is also the ground state of the Hamiltonian. To analyze this, we proceed in analogy to previous works \cite{PhysRevX.10.031034,PhysRevB.103.205414,christos2022correlated,7z4z-vlj8} and rewrite $H_V$ as
\begin{equation}
    H_V = \frac{1}{2N} \sum_{\vec{q}} V_{\vec{q}} \Delta \rho_{\vec{q}} \Delta \rho_{-\vec{q}} + \frac{1}{N} \sum_{\vec{G} \in \text{RL}} V_{\vec{G}} \lambda(-\vec{G}) \rho_{\vec{G}} - \frac{1}{2N} \sum_{\vec{G} \in \text{RL}} V_{\vec{G}} |\lambda(\vec{G})|^2 := \widetilde{H}_V + \frac{1}{N} \sum_{\vec{G} \in \text{RL}} V_{\vec{G}} \lambda(-\vec{G}) \rho_{\vec{G}}
\end{equation}
where we defined the shifted density operator $\Delta \rho_{\vec{q}} = \rho_{\vec{q}} - \delta_{\vec{q} \in \text{RL}} \lambda(\vec{q})$. Due to \equref{HermiticityConstraint}, we see that $\Delta \rho_{-\vec{q}} = (\Delta \rho_{\vec{q}})^\dagger$ such that $\frac{1}{2N} V_{\vec{q}} \sum_{\vec{q}} \Delta \rho_{\vec{q}} \Delta \rho_{-\vec{q}}$ is positive semi-definite and $\ket{\text{FP}(\mathcal{S})}$ is a ground state of $\widetilde{H}_V$. However, $\Delta H_V = H_V - \widetilde{H}_V$ can in general increase its energy such that $\ket{\text{FP}(\mathcal{S})}$ ceases to be the ground state. A scenario where we can guarantee that this is not the case is the regime where the ``flat-metric condition'' (FMC),
\begin{equation}
    \Lambda_\eta(\vec{k},\vec{G}) = f(\vec{G}), \label{FMCDef}
\end{equation}
holds. In \appref{App:FMCSU3}, we will check the validity of this assumption and quantify deviations from it. If \equref{FMCDef} holds, we can write
\begin{equation}
    \Delta H_V \equiv \frac{1}{N} \sum_{\vec{G} \in \text{RL}} V_{\vec{G}} \lambda(-\vec{G}) \rho_{\vec{G}} = \left[\sum_{\vec{G} \in \text{RL}} V_{\vec{G}} \lambda(-\vec{G}) f(\vec{G}) \right] \hat{\nu} , \quad \hat{\nu} = \frac{1}{N}\sum_{\vec{k},\eta,s} c^\dagger_{\vec{k},\eta,s} c^\pdagger_{\vec{k},\eta,s},
\end{equation}
which is simply a constant at fixed filling fraction. Interestingly, this already captures a significant fraction of the phases in \figref{Fig2:HFphasediagram} of the main text: it includes the states VP$_{\uparrow}$ and VP$_{\uparrow\downarrow}$, e.g., $\mathcal{S}=\{(1,\uparrow)\}$ and $\mathcal{S}=\{(1,\uparrow),(2,\downarrow)\}$ as representatives of the order-parameter manifold for $\nu=1$ and $\nu=2$, respectively, and the SP state at $\nu=3$ via, e.g., $\mathcal{S} = \{(1,\uparrow),(2,\uparrow),(3,\downarrow)\}$ as a representative. This provides some understanding of the region of small twist angle $\theta$ for AB stacking and for intermediate $\theta$ for AA stacking. However, it does not explain the emergence of the IVC phases in the other regimes, in particular at small $\theta$ and $\epsilon_r$ which we expect to correspond to the strong-coupling regime. In the remainder of this appendix, we will identify the driving force behind these additional orders. 

\subsection{U(6) symmetric limit}
To this end, let us first further fine-tune the Hamiltonian before approaching again the more realistic parameter regimes perturbatively. Since they are diagonal in valley (and trivial in the spin) index, the form factors have three independent (in general complex-valued) components. It will be convenient to parametrize them as
\begin{equation}
    \Lambda_{\eta}(\vec{k},\vec{q}) = \left[\frac{1}{\sqrt{3}}\lambda_0(\vec{k},\vec{q}) \begin{pmatrix} 1 \\ 1 \\ 1 \end{pmatrix} + \frac{1}{\sqrt{2}}\lambda_x(\vec{k},\vec{q}) \begin{pmatrix} 1 \\ -1 \\ 0 \end{pmatrix} + \frac{1}{\sqrt{6}}\lambda_y(\vec{k},\vec{q}) \begin{pmatrix} 1 \\ 1 \\ -2 \end{pmatrix}\right]_\eta. \label{FormFactorRewriting}
\end{equation}
Then, $\lambda_0$ transforms under the trivial representation ($A$) of $C_{3z}$ while $(\lambda_x,\lambda_y)$ transforms under the vector representation ($E$) of it. For AA stacking, $\lambda_0$, $\lambda_{x,y}$ are in general complex while the presence of the $C_{2z}\Theta$ symmetry makes them real for AB stacking. 

Let us first assume that $|\lambda_{x,y}| \ll |\lambda_z|$ such that we can set $\lambda_{x,y} = 0$. Then, the form factors are independent of $\eta$ and the model is fully U(3) symmetric in valley space or U(6) symmetric in the combined valley-spin space,
\begin{equation}
    \textrm{U}(6):\quad c_{\vec{k},\eta,s} \quad \rightarrow \quad (U)_{\eta,s;\eta',s'} c_{\vec{k},\eta',s'},\qquad U \in \text{U}(6). \label{U6Symmetry}
\end{equation}
We emphasize that this symmetry can still be present, even away from the flat-band limit.

To describe the conditions for this symmetry for the dispersion in an analogous way, we write the non-interacting part of the Hamiltonian as
\begin{subequations}
\begin{equation}
    H_0 = \sum_{\vec{k}} \sum_{\eta,s} \epsilon_{\vec{k},\eta} c^\dagger_{\vec{k},\eta,s} c^\pdagger_{\vec{k},\eta,s},
\end{equation}
and parametrize
\begin{equation}
    \epsilon_{\vec{k},\eta} = \left[\frac{1}{\sqrt{3}}\epsilon_0(\vec{k}) \begin{pmatrix} 1 \\ 1 \\ 1 \end{pmatrix} + \frac{1}{\sqrt{2}}\epsilon_x(\vec{k}) \begin{pmatrix} 1 \\ -1 \\ 0 \end{pmatrix} + \frac{1}{\sqrt{6}}\epsilon_y(\vec{k}) \begin{pmatrix} 1 \\ 1 \\ -2 \end{pmatrix}\right]_\eta,
\end{equation}\label{NoninterActingPart}\end{subequations}
where, again, $\epsilon_0$ is invariant under $C_{3z}$ while $(\epsilon_x,\epsilon_y)$ transform under the two-dimensional representation $E$. Clearly, the U(6) symmetry in \equref{U6Symmetry} is present for $\epsilon_{x,y}=0$.  

Back to the flat-band limit, where the flavor-polarized states in \equref{FPStates} become the exact ground states. In the presence of the aforementioned U(6) symmetry, the same must hold for the family of states given by
\begin{equation}
    \ket{\text{FP}(\mathcal{S};U)} := \prod_{(s,\eta)\in \mathcal{S}} \prod_{\vec{k}} [U_{s,\eta,s',\eta'}c_{\vec{k},\eta',s'} ]^\dagger \ket{0}, \label{RotatedFPStates}
\end{equation}
where summation over repeated indices is implied and $U$ is one of the transformation matrices of Eq.~\eqref{U6Symmetry}. 
This now also contains IVC phases. Since all such states are product states, we can also characterize them by the correlator matrix
\begin{equation}
    (P_{\vec{k}})_{s,\eta;s',\eta'} := \braket{c^\dagger_{\vec{k},s,\eta} c^\pdagger_{\vec{k},s',\eta'}}.\label{correlator}
\end{equation}
For the states in \equref{RotatedFPStates}, we have $P_{\vec{k}} = P$ and all eigenvalues are either $0$ or $1$ such that $P^2=P$. Under the U(6) transformation in \equref{U6Symmetry}, it holds
\begin{equation}
    \textrm{U}(6): \quad P \quad \rightarrow \quad U^\dagger P U.
\end{equation}
This shows that in the U(6) symmetric, flat-band and flat-metric limit, any product state with a momentum independent $P$ obeying $P^2 = P$ and the filling constraint $\text{tr} (P) = \nu$ is an exact ground state at integer filling $\nu$. While this defines a continuum of possible ground states, we can now combine this with the general symmetry constraints from \appref{GLanal} to arrive at a finite discrete set of physical candidate orders.

For simplicity, we focus on filling $\nu=2$ and assume that spin is unpolarized, i.e., that both spin-up and spin-down are occupied in exactly the same way. Alternatively, our discussion applies to $\nu=1$ as well with the additional assumption that only one spin flavor is occupied. In both cases, this will allow us to focus on the valley-flavor space, i.e., only study the U(3) subgroup in valley space and specify $P$ as $3 \times 3$ matrices in valley space.
\begin{enumerate}
    \item Let us start with the valley-diagonal states, i.e., those without any form of intervalley coherence and where the $[\textrm{U}(1)]_\textrm{V}^3$ symmetry is completely preserved. Since all eigenvalues can only be $1$ or $0$, only
    \begin{equation}
        P_{\text{VP}} = \begin{pmatrix}
            0 & 0 & 0 \\
            0 & 0 & 0 \\
            0 & 0 & 1 
        \end{pmatrix}
    \end{equation}
    and those obtained by permutation of the three valleys are consistent with $\text{tr}(P) =1$. As already anticipated by the subscript, this is simply the valley-polarized state in the first line of \tableref{SymmetriesOfStates}. It is also part of the flavor polarized states in \equref{FPStates}. To clarify the meaning of this state upon including spin: for $\nu=1$, it corresponds to occupying one spin and one valley, whereas both spin-up and spin-down are occupied in one valley in the case of a filling of $\nu=2$.
    
    \item We next continue with the IVC$_+^3$. This state breaks $[\textrm{U}(1)]_{\textrm{V}}^3$ down to the $\textrm{U}(1)_\textrm{C}$ associated with total charge conservation. It is further characterized by the existence of a $[\textrm{U}(1)]_\textrm{V}^3$ transformation that brings the ground state to a form [$\varphi_{ij}=0$ in \equref{DescriptionOfStates}] such that it is explicitly invariant under $C_{3z}$, without having to accompany it by a $[\textrm{U}(1)]_\textrm{V}^3$ transformation. In this form, it holds
    \begin{equation}
        P_{\text{IVC}_+^3} = \frac{1}{3}  \begin{pmatrix}
            1 & 1 & 1 \\ 1 & 1 & 1 \\ 1 & 1 & 1
        \end{pmatrix}. \label{PIVC3pForm}
    \end{equation}
    The easiest way to arrive at this form is to note that $\text{tr}(P) =1$ and $P^2 = P$ necessitate that $P$ be of the form $P=v v^T$ with some three-component, normalized column vector $v$. Being invariant under $C_{3z}$ then immediately implies $v=e^{i\varphi} (1,1,1)^T/\sqrt{3}$, leading to \equref{PIVC3pForm}. 

    \item For the IVC$_-^3$ state, the main difference is that there is no $[\textrm{U}(1)]_\textrm{V}^3$ transformation such that the state is explicitly invariant under $C_{3z}$; instead, $C_{3z}$ always has to be accompanied by some $[\textrm{U}(1)]^3_\textrm{V}$ rotation. Let us take $\varphi_{12}=\pi$, $\varphi_{23}=\varphi_{31}=0$ in \equref{DescriptionOfStates} without loss of generality. This means that $C_{3z}$ has to be accompanied by $U = \text{diag}(1,-1,1)$. For the same reason as above, we must have $P=v v^T$ and we now obtain the constraint
    \begin{equation}
        U^T C_{3z}^T v = v \quad \Leftrightarrow \quad \begin{cases}
            v_3 = v_1 \\ -v_1=v_2 \\
            v_2 = v_3
        \end{cases}
    \end{equation}
    which implies $v=0$, which cannot be normalized. This shows that the IVC$_-^3$ state cannot be realized at this filling. Alternatively, this can also be seen by diagonalizing $-\vec{\phi}$ with $\vec{\phi}_{\eta,\eta'}$ as given in  \equref{DescriptionOfStates} for the IVC$_-^3$. One indeed finds a doubly degenerate ground state. This explains why $P\neq v v^T$. Below we state the form of $P$ for this state since it does become important at a filling of $\nu=2$ if spin is polarized; this will be discussed in \secref{DeviationsFromFMC}.

    \item We continue with NIVC$^1$, where only two valleys develop coherence and, as expected by symmetry, the occupations in the three valleys are different. In terms of symmetries, $[\text{U}(1)]_\textrm{V}^3$ is now broken down to $\text{U}(1)_\text{C} \times \text{U}(1)_\textrm{V}$ where $\text{U}(1)_\text{V}$ corresponds to charge conservation in the uncoupled valley while $\text{U}(1)_\text{C}$ just represents conservation of the total charge. Naturally, this automatically breaks $C_{3z}$.
    
    It is straightforward to see that there is a one-parameter family of states that are all degenerate in the U(3)-symmetric limit:
    \begin{equation}
        P_{\text{NIVC}^1} = \frac{1}{1+\alpha^2}\begin{pmatrix}
            \alpha^2 & \alpha & 0 \\ \alpha & 1 & 0 \\ 0 & 0 & 0
        \end{pmatrix}.
    \end{equation}
    This means that one valley remains unoccupied while two have unequal occupations but develop coherence. Out of these two coherent superpositions only one will be occupied. As we will see below, energetics beyond the U(3)-symmetric limit will determine the value of $\alpha$. We note that $\alpha\rightarrow 0$ (or $\alpha\rightarrow \infty)$ reproduces the VP state.
    \item Finally, we discuss NIVC$^3$. Just like with IVC$_{\pm}^3$, this phase also breaks $[\text{U}(1)]_\textrm{V}^3$ down to $\text{U}(1)_\text{C}$. The difference, however, is that there is no residual $C_{3z}$ symmetry and, thus, the ground-state occupations in the three valleys is in general different. The only additional symmetry constraint is that there is always a $[\text{U}(1)]_\textrm{V}^3$ transformation after which the state is explicitly invariant under time-reversal $\Theta$. This implies that $v$ can be chosen real and we parametrize $v=(\alpha,1,\beta)/\sqrt{\alpha^2+1+\beta^2}$. The resulting 
    \begin{equation}
        P_{\text{NIVC}^3} = \frac{1}{\alpha^2+1+\beta^2}\begin{pmatrix}
            \alpha^2 & \alpha  & \alpha \beta \\ \alpha  & 1 & \beta \\ \alpha \beta & \beta & \beta^2
        \end{pmatrix} \label{PNIVC3pm}
    \end{equation}
    now depends on two real parameters $\alpha,\beta\in\mathbbm{R}$. The limit $\beta \rightarrow 0$ (or $\alpha\rightarrow 0$ for that matter) corresponds to NIVC$^1$ and further setting $\alpha\rightarrow 0,\infty$ (or $\beta\rightarrow 0,\infty$) leads to the VP state. Furthermore, setting $\alpha=\beta=1$ reproduces the IVC$_+^3$ state. 

    We note that the form of $\phi$ from the Ginzburg-Landau analysis and as stated in \equref{DescriptionOfStates} of the main text is more restricted for this state: two of the three off-diagonal elements have the same magnitude. This is also the reason why we indicated that the state is invariant under $C_{2x}$ in \tableref{SymmetriesOfStates}. In \equref{PNIVC3pm}, this would correspond to $\alpha =1$ (or $\beta=1$). We kept the analytics more general since any state of the form of \equref{PNIVC3pm} is part of the U(6) manifold, allowing us to explore a larger set of wavefunctions. Besides, we are here trying to capture ground-state and, thus, zero-temperature properties, instead of focusing on the vicinity of the critical temperature, where the Ginzburg-Landau analysis applies.
\end{enumerate}

\subsection{U(6)-symmetry-breaking form factors as perturbation}
With those four (families of) candidate orders at hand, which are all exactly degenerate in the U(6)-symmetric limit, we now study the impact of the terms $\lambda_{x,y}$ in the form factors in \equref{FormFactorRewriting} on relative energetics. Within leading-order perturbation theory, we simply need to evaluate the expectation value of the states with respect to the perturbation. Due to their product-state nature, this is equivalent to evaluating the Hartree-Fock (HF) ground-state energies cf.~\equref{HFener}.

In \tableref{FormFactorPerturbations}, we list the Fock contributions for the different candidate orders. To obtain compact expressions, we defined
\begin{equation}
    \braket{\dots} :=  \frac{1}{2N} \sum_{\vec{k}} \sum_{\vec{q}} V_{\vec{q}} \dots \,. \label{DefinitionOfExpVal}
\end{equation}
To arrive at these expressions, we have made use of the orthogonality of the basis functions, e.g., $\braket{\lambda^*_0\lambda_x}=0$.
Note that there is no need to include the Hartree contribution since it is the same for every state and given by
\begin{equation}
    E_{\text{H}} = \frac{1}{6N} \sum_{\vec{G}} V_{\vec{G}} |\sum_{\vec{k}}\lambda_0(\vec{k},\vec{G})|^2 = \frac{N}{2}\sum_{\vec{G}} V_{\vec{G}} |f(\vec{G})|^2.
\end{equation}
The reason is that we still assume the FMC (\ref{FMCDef}) such that $\lambda_{x,y}(\vec{k},\vec{G}) = 0$. We will relax this assumption in the next subsection. 

\begin{table}[tb]
\begin{center}
\caption{{\bf{Fock energies for the different candidate orders.}} \justifying We take the form factors $\lambda_{x,y}$ as a perturbation and still assume the FMC. The expectation value $\braket{\dots}$ is defined in \equref{DefinitionOfExpVal}. Although NIVC$^1$ and NIVC$^3_\pm$ have free parameters, their optimal energy is reached for fine-tuned points that correspond to the VP state, see last column. The energies refer to a filling of $\nu=1$. Multiplying them by $2$ yields the energies of the corresponding spin-unpolarized states at $\nu=2$.}
\label{FormFactorPerturbations}\begin{ruledtabular}\begin{tabular}{ccc}
State & Fock energy & Optimal parameters   \\ \hline
VP & $-\frac{1}{3} \braket{|\lambda_0|^2+|\lambda_x|^2+|\lambda_y|^2}$ & ---  \\
IVC$_+^3$ & $-\frac{1}{3} \braket{|\lambda_0|^2}$ & ---   \\
NIVC$^1$ & $-\frac{1}{3} \braket{|\lambda_0|^2} -\frac{1}{12}\braket{|\lambda_x|^2+|\lambda_y|^2}\left[1 + 3\frac{(\alpha^2-1)^2}{(\alpha^2+1)^2}\right]$ & $\alpha=0,\infty$ (VP)    \\
NIVC$^3$ & $-\frac{1}{3} \braket{|\lambda_0|^2} -\frac{1}{12}\braket{|\lambda_x|^2+|\lambda_y|^2}\left[\frac{(\alpha^2+1-2\beta^2)^2}{(\alpha^2+1+\beta^2)^2} + 3\frac{(\alpha^2-1)^2}{(\alpha^2+1+\beta^2)^2}\right]$ & $(\alpha,\beta) = (0,0),(\infty,0),(0,\infty)$ (VP)   \\
 \end{tabular}
\end{ruledtabular}
\end{center}
\end{table}

We can see that if $\lambda_{x,y}$ are the dominant perturbation to the U(6)-symmetric, flat-band, and flat-metric limit, the VP will be favored. This is not surprising since we established earlier that the VP state is the (or at least one of the) exact ground state (states) in the flat-band and flat-metric limit. The current analysis, however, shows that the other candidate orders are not also exactly degenerate ground states in this limit. It will further form the basis for the following investigations.

\subsection{Deviations from the FMC}\label{DeviationsFromFMC}
Next, we treat deviations from the FMC (\ref{FMCDef}) as a perturbation. To first order, as before, this corresponds to evaluating the HF energy. In fact, since the FMC did not play a role in our analysis above for the Fock term, it will only affect the Hartree energies. We list them in \tableref{HartreeEnergies}, where we defined
\begin{equation}
    \braket{ |\lambda_j|^2 }_{G} := \frac{1}{2N} \sum_{\vec{G}} V_{\vec{G}} |\sum_{\vec{k}} \lambda_j(\vec{k},\vec{G})|^2. \label{ExpectGDef}
\end{equation}

Interestingly, deviations from the FMC lead to a rather different energetics: now, the VP receives the largest energy penalty. Among the remaining IVC states, we see that the IVC$_+^3$ is now favored (the optimal NIVC$^3_\pm$ becomes the IVC$_+^3$ while the energy of the optimal NIVC$^1$ phase is higher). Note that this perturbation also induces significant momentum dependence in the IVC$_+^3$ state. Instead of discussing this on the level of perturbation theory, we will study it variationally in \secref{MomentumDependentIVC} below, where it is found to induce significant $\vec{k}$ dependencies due to the complex nature of the form factors for AA stacking (in agreement with our numerics).
Note that the energies in \tableref{HartreeEnergies} become degenerate if $\lambda_{x,y}=0$. This means that violation of the FMC has to also break U(6) in order the uniquely select the ground state. In fact, we will see in \secref{App:FMCSU3} below that the deviations from \equref{FMCMainText} primarily come from the $\eta$ dependence in the concrete model of twisted $\textrm{SnSe}_2$ which we focus our explicit calculations on.

Taken together, we expect that either the VP state or the IVC$_+^3$ phase is favored in the flat-band limit at $\nu=1$; the VP (IVC$_+^3$) state is expected to be the ground state if deviations from U(3) symmetric form factors are dominant over (dominated by) the deviations from the FMC. These conclusions are in line with the small-$\theta$/small-$\epsilon_r$ limit of our Hartee-Fock phase diagram in \figref{Fig2:HFphasediagram}, where exactly these two phases appear. For larger $\theta$ and/or $\epsilon_r$, the impact of the dispersion cannot be neglected. We take it into account in the next subsection.

Before that, however, let us revisit the IVC$^3_-$ state. While it cannot be realized at $\nu=1$ (nor at $\nu=2$ if we assume that there is no spin polarization), it is part of the U(6) manifold at filling $\nu=2$ if we fill only one spin flavor. The associated $P$, again only shown in valley space, reads as
\begin{equation}
    P_{\text{IVC}^3_-} = \frac{1}{3}  \begin{pmatrix}
            2 & -1 & 1 \\ -1 & 2 & 1 \\ 1 & 1 & 2
        \end{pmatrix}.
\end{equation}
Let us now study the impact of deviations from the FMC by evaluating the Hartree energy. One finds $\frac{4}{3}\braket{|\lambda_0|^2}_G$, i.e., these deviations favor the spin-unpolarized versions of IVC$^3_-$ over the VP, NIVC$^1$, and NIVC$^3_{\pm}$ state at $\nu=2$ (cf.~\tableref{HartreeEnergies}). At the same time, the Hartree energy is identical to the spin-unpolarized IVC$^3_+$ phase at $\nu=2$. So the energetic splitting between IVC$^3_-$ and IVC$^3_+$ comes from deviations of the form factors breaking U(6) symmetry. As before, we probe this within first-order perturbation theory, which is equivalent to evaluating the Fock energy. We find for the IVC$^3_-$ at $\nu=2$ a Fock energy of $-\frac{2}{3} \braket{|\lambda_0|^2|}-\frac{1}{3}\braket{|\lambda_x|^2+|\lambda_y|^2}$. Interestingly, the IVC$^3_-$ state can indeed benefit from the form factor deviations, as opposed to the (at $\nu=2$ spin-unpolarized) IVC$^3_+$. This energetic benefit is still lower than that of the spin-unpolarized VP state at $\nu=2$, for which the Fock energy reads as $-\frac{2}{3} \braket{|\lambda_0|^2+|\lambda_x|^2+|\lambda_y|^2}$, cf.~\tableref{FormFactorPerturbations}. As such, the conclusions above for $\nu=1$ carry over to $\nu=2$, upon replacing IVC$^3_+$ by IVC$^3_-$; this agrees well with the numerical results in \figref{Fig2:HFphasediagram}.

\begin{table}[tb]
\begin{center}
\caption{{\bf Hartree energies for the different candidate orders.} \justifying We take the deviation from the FMC as a perturbation. See \equref{ExpectGDef} for the definition of the expectation value. The energies refer to a filling of $\nu=1$ (spin is polarized). Multiplying them by $4$ yields the energies of the respective spin-unpolarized states at $\nu=2$.}
\label{HartreeEnergies}\begin{ruledtabular}\begin{tabular}{ccc}
State & Hartree energy & Optimal parameters   \\ \hline
VP & $\frac{1}{3}[\braket{|\lambda_0|^2}_G + \braket{|\lambda_x|^2}_G +\braket{|\lambda_y|^2}_G]$ & ---  \\
IVC$_+^3$ & $\frac{1}{3}\braket{|\lambda_0|^2}_G $ & ---   \\
NIVC$^1$ & $\frac{1}{3} \braket{|\lambda_0|^2}_G +\frac{1}{12}(\braket{|\lambda_x|^2}_G+\braket{|\lambda_y|^2}_G)\left[1 + 3\frac{(\alpha^2-1)^2}{(\alpha^2+1)^2}\right]$ & $\alpha=1$ (NIVC$^1$)    \\
NIVC$^3$ & $\frac{1}{3} \braket{|\lambda_0|^2}_G +\frac{1}{12}[\braket{|\lambda_x|^2}_G+\braket{|\lambda_y|^2}_G]\left[\frac{(\alpha^2+1-2\beta^2)^2}{(\alpha^2+1+\beta^2)^2} + 3\frac{(\alpha^2-1)^2}{(\alpha^2+1+\beta^2)^2}\right]$ & $(\alpha,\beta) = (1,1)$ (IVC$_+^3$)   \\
 \end{tabular}
\end{ruledtabular}
\end{center}
\end{table}

\subsection{Dispersion as a perturbation}
Finally, we will study the non-interacting part of the Hamiltonian in \equref{NoninterActingPart} as a perturbation. Using $C_{3z}$ rotational symmetry, we find that the first-order contribution to the energy is $\frac{\nu}{\sqrt{3}} \sum_{\vec{k}\in\textrm{mBZ}} \epsilon_{0}(\vec{k})$ for all states. Consequently, we have to go to second order in $H_0$ to obtain a splitting between the different candidate orders. Instead of explicitly evaluating it, we use the criterion that a quadratic term in the Hamiltonian, $\sum_{\vec{k}\in\textrm{mBZ}} c^\dagger_{\vec{k}} M_{\vec{k}} c^\pdagger_{\vec{k}}$, contributes in second-order perturbation theory and, thus, lowers the energy if it does not commute with $P$ (see, e.g., Appendix SI.F.2 in \cite{christos2022correlated}). In our case, the different $M_{\vec{k}}$ are
\begin{equation}
    \frac{1}{\sqrt{3}} \epsilon_0(\vec{k})\mathbbm{1}_{3\times3}, \qquad \frac{1}{\sqrt{2}} \epsilon_x(\vec{k}) \begin{pmatrix}
        1 & 0 & 0 \\ 0 & -1 & 0 \\ 0 & 0 & 0
    \end{pmatrix}, \qquad \frac{1}{\sqrt{6}} \epsilon_y(\vec{k}) \begin{pmatrix}
        1 & 0 & 0 \\ 0 & 1 & 0 \\ 0 & 0 & -2
    \end{pmatrix}\label{disper}.
\end{equation}
Clearly, the first term commutes with any $P$ and, hence, does not contribute to the energy. For the remaining two we find:
\begin{enumerate}
    \item VP: its $P$ commutes with both terms. As such, the VP state cannot benefit energetically from the kinetic energy. This is intuitively clear since the kinetic terms do not scatter between the different valleys (or spins), so there is no accessible virtual state with exactly one valley occupied.
    \item IVC$_+^3$ does not commute with either of those two terms and, hence, lowers its energy due to these superexchange processes. 
    \item NIVC$^1$ lowers its energy too, however, only due to the $\epsilon_x(\vec{k})$ term since it commutes with the matrix multiplying $\epsilon_y(\vec{k})$. This asymmetry comes from the fact that the NIVC$^1$ state breaks $C_{3z}$ rotational symmetry.
    \item Finally, the low-symmetry NIVC$^3_\pm$ state does not commute with either of the two terms, both lowering its energy.
\end{enumerate}
In summary, we see that the dispersion will favor IVC states over the VP state. This is also in line with our HF phase diagram in \figref{Fig2:HFphasediagram}.

\subsection{Momentum dependence of IVC$^3_{\pm}$ order}\label{MomentumDependentIVC}
Finally, we consider the relevance of momentum dependence in the IVC$^3_{\pm}$ order parameter, using a variational approach. Focusing on the IVC$^3_{+}$ order (at $\nu=1$ with spin polarization), we parametrize the correlator matrix as \begin{align}
\ P_{\vec{k},\textrm{IVC}^3_+} \equiv \frac{1}{3} \left ( \begin{array}{cccc} 1 & e^{i\theta_1(\vec{k})} & e^{-i[\theta_1(\vec{k})+\theta_2(\vec{k})]}\\
             e^{-i\theta_1(\vec{k})} & 1 &e^{i\theta_2(\vec{k})} \\
             e^{i[\theta_1(\vec{k})+\theta_1(\vec{k})]} &e^{-i\theta_2(\vec{k})}&1\\
             \end{array} \right).
\end{align} 
We list an analytical expression for the HF energy of this ansatz in \tableref{momdepFormFactorPerturbations}. Here, we defined
\begin{subequations}
  \begin{align}
\begin{split}
\Delta_{1 (2),\vec{k,q}}\equiv\theta_{1 (2),\vec{k+q}}-\theta_{1 (2),\vec{k}}, \  \ \  \Delta_{3,\vec{k,q}}\equiv\Delta_{1,\vec{k,q}}+\Delta_{2,\vec{k,q}},
\end{split}\\
\begin{split}
\left(  \begin{array}{cccc}\lambda_1(\vec{k,q})\\\lambda_2(\vec{k,q})\\\lambda_3(\vec{k,q})\\ \end{array}\right) \equiv  \left ( \begin{array}{cccc} \frac{1}{\sqrt{3}} & \frac{1}{\sqrt{2}} & \frac{1}{\sqrt{6}}\\
            \frac{1}{\sqrt{3}} & -\frac{1}{\sqrt{2}} & \frac{1}{\sqrt{6}}\\
             \frac{1}{\sqrt{3}} & 0 & -\frac{2}{\sqrt{6}}\\
             \end{array} \right) \left(  \begin{array}{cccc}\lambda_0(\vec{k,q})\\\lambda_x(\vec{k,q})\\\lambda_y(\vec{k,q})\\ \end{array}\right)
\end{split}\end{align}\label{addparameter}\end{subequations}
For the U(6) symmetric limit, we see that the energy of the state is lower bounded to $\frac{1}{3} \braket{|\lambda_0|^2}_{G}-\frac{1}{3}\braket{|\lambda_0|^2}$, where the equality is satisfied when $\Delta_{j,\vec{k,q}}=0$. This corresponds to momentum independent IVC$^3_+$. Taking the deviation from the FMC into account does not affect the form of the energy of this state. However, violating the U(6) symmetry affects the Fock energy of the state. Using the Cauchy inequality, we see that the total energy is upper bounded to $\frac{1}{3} \braket{|\lambda_0|^2}_{G}-\frac{1}{3}\braket{|\lambda_0|^2}$, where the equality is satisfied when $\Delta_{j,\vec{k,q}} = -\textrm{Arg}[\lambda^*_j\lambda_{j+1}]$. This shows that $\Delta_{j,\vec{k,q}}\neq0$ for AA stacking where there is in general no gauge with entirely real $\lambda_j$; this corresponds to momentum dependent IVC$^3_{+}$ order. However, for AB stacking, where $C_{2z}\Theta$ can be used to make $\lambda_j$ real, this leads to $\Delta_{j,\vec{k,q}}=0$, which corresponds to momentum independent IVC$^3_{+}$ order.

In addition, we see that the competition between the Hartree and Fock contributions determines whether the VP or IVC$^3_{+}$ state becomes the ground state. We point out that when we only consider the Fock contribution without U(6) symmetry, our variational argument here shows that VP order has lower energy than both momentum independent and dependent IVC$^3_+$ order, which we have also checked agrees with our numerics when we artificially set the Hartree contribution to zero.

In conclusion, based on our analytics, we find that it is the Hartree energy that can favor an IVC$^3_+$ over the VP state. We furthermore see that the complex nature of the form factors (which cannot be removed by a gauge transformation for AA stacking) leads to momentum dependencies in the IVC$^3_+$ order if U(6)-symmetry breaking contributions are taken into account in the Fock term. We have checked that these conclusions agree with our numerical unrestricted HF calculations.

\begin{table}[tb]
\begin{center}
\caption{{\bf{Energy of momentum-dependent IVC$^3_+$ order}}. \justifying We track the energy of momentum dependent IVC$^3_+$ order for each possible perturbation $\in\{\emptyset, \ \textrm{FMC}, \ \text{U}(6)\}$, where $\emptyset$ corresponds to U(6) symmetric limit. That is, the first line shows the variational energy for the U(6)-symmetric limit, which has the same form irrespective of whether the FMC is obeyed or not. When the U(6) symmetry is broken, there are additional contributions as shown in the second line. See \equref{DefinitionOfExpVal}, \equref{ExpectGDef} for the definition of the expectation values and \equref{addparameter} for the parameters.For notational simplicity, we suppress the momentum dependence of the form factors here.}
\label{momdepFormFactorPerturbations}\begin{ruledtabular}\begin{tabular}{ccc}
Perturbation & Hartree energy + Fock Energy & Optimal parameters   \\ \hline
$\emptyset$, FMC & $\frac{1}{3} \braket{|\lambda_0|^2}_{G}-\frac{2}{27}\braket{|\lambda_0|^2(\frac{3}{2}+\textrm{cos}[\Delta_{1,\vec{k,q}}]+\textrm{cos}[\Delta_{2,\vec{k,q}}]+\textrm{cos}[\Delta_{3,\vec{k,q}}]}$ & $\Delta_{j,\vec{k,q}}=0$  \\
U(6) & $\frac{1}{3} \braket{|\lambda_0|^2}_{G}-\frac{1}{9}\braket{|\lambda_1|^2+|\lambda_2|^2+|\lambda_3|^2+2\textrm{Re}[\lambda^*_1\lambda_2e^{i\Delta_{1,\vec{k,q}}}+\lambda^*_2\lambda_3e^{i\Delta_{2,\vec{k,q}}}+\lambda^*_3\lambda_1e^{i\Delta_{3,\vec{k,q}}}]}$ & $\Delta_{j,\vec{k,q}}=-\textrm{Arg}[\lambda^*_j\lambda_{j+1}]$  \\
 \end{tabular}
\end{ruledtabular}
\end{center}
\end{table}

\section{HF calculation}
\label{appenHFeq}
In this section, we describe the HF calculation in detail. We perform the calculation for the lowest single active band per spin and valley for computational efficiency and project the dual-gated Coulomb interaction into this band. For given moir\'e momentum $\vec{k}\in\textrm{mBZ}$, spin $s$ and valley $\eta$ degrees of freedom, we first start with the single particle continuum Hamiltonian written in band, spin and valley space as $[H_{\textrm{cont}}(\vec{k})]_{s,\eta;s' \eta'} = \epsilon_{s,\eta}(\vec{k})\delta_{s,s'}\delta_{\eta,\eta'}$, with single particle dispersion of the active band $\epsilon_{s,\eta}(\vec{k})$. For the interaction, we use \equref{HVInteraction} with a double-gate screened potential $V_{\vec{q}} = \frac{e^2}{2\epsilon_r \epsilon_0 |\vec{q}|}\textrm{tanh}(|\vec{q}| d)$. Here, we defined the dielectric constant $\epsilon_r$, electron charge $e$ and screening length $d$. In our work, we fix $d=40\ \textrm{nm}$ and set $\epsilon_r$ as a tunable parameter, which determines the interaction strength as $V_{\vec{q}}\propto1/\epsilon_{r}$. Note that $V_{\vec{q}}=V_{-\vec{q}}$ and $V_{\vec{q}}>0$.

After a mean field decoupling process of \equref{HVInteraction}, the HF Hamiltonian $H_{\textrm{HF}}(\vec{k},P_{\vec{k}})$ can be represented as a sum of Hartree and Fock contributions, $H_{\textrm{HF}}(\vec{k},P_{\vec{k}})=H_{\textrm{Hartree}}(\vec{k},P_{\vec{k}})+H_{\textrm{Fock}}(\vec{k},P_{\vec{k}})$, with
\begin{subequations}
\begin{align}
\begin{split}
H_{\textrm{Hartree}}(\vec{k},P_{\vec{k}}) = \frac{1}{N}\sum_{\vec{G}\in\textrm{RL}}V_{\vec{G}}\Lambda^{\dagger}(\vec{k,G})\sum_{\vec{k}'\in\textrm{mBZ}}\textrm{tr}\left[(P_{\vec{k}'}-P^0_{\vec{k}'})\Lambda^T(\vec{k}',\vec{G})\right]\nonumber
\end{split} \\
\begin{split}
H_{\textrm{Fock}}(\vec{k},P_{\vec{k}})=-\frac{1}{N}\sum_{\vec{q}\in\mathbbm{R}^2}V_{\vec{q}}\Lambda^{\dagger}(\vec{k,q})(P_{\vec{k}+\vec{q}}-P^{0}_{\vec{k}+\vec{q}})^T\Lambda(\vec{k,q})
\end{split}\nonumber, 
\end{align}
\end{subequations}
and the correlator matrix $P_{\vec{k}}$ defined in \equref{correlator}. In our work, we only take (moiré) translation invariant solutions into account, i.e., for given two moir\'e momenta $\vec{k}, \vec{k}' \in\textrm{mBZ}$, we impose the constraint $\langle c^{\dagger}_{\vec{k},s,\eta}c^{\dagger}_{\vec{k}',s',\eta'} \rangle \sim \delta_{\vec{k},\vec{k}'}$.

To avoid double counting of interaction effects, we subtract the reference scheme $P^0_{\vec{k}}$ from the correlator $P_{\vec{k}}$. In our work, we use the 'average' scheme \cite{parker2021field}, where $P^0_{\vec{k}}=\mathbbm{1}/2$ for active bands, $P^0_{\vec{k}}=\mathbbm{1}$ for occupied remote bands, and $P^0_{\vec{k}}=0$ for unoccupied remote bands. We point out that choosing the empty system as a reference scheme (as we did in the strong-coupling analysis) does not qualitatively change our HF results; it only shifts the energy of each state by a constant compared to the energy in the 'average' scheme.

For each integer filling $\nu$, we start with several random ans\"atze of $P_{\vec{k}}$, perform an iteration with the Hamiltonian $H_{\textrm{cont}}(\vec{k})+H_{\textrm{HF}}(\vec{k}, P_{\vec{k}})$ until convergence of $P_{\vec{k}}$ is reached, and pick the $P_{\vec{k}}$ which gives the lowest energy $E_{\textrm{HF}}(P_{\vec{k}})$. In addition, to take the flat band limit ($\epsilon_r\rightarrow 0$) in our numerics, we perform a HF calculation with the Hamiltonian $H_{\textrm{HF}}(\vec{k}, P_{\vec{k}})$ only, where the single particle band dispersion is neglected ($H_{\textrm{cont}}(\vec{k})=0$). For given correlator $P_{\vec{k}}$, the corresponding energy $E_{\textrm{HF}}$ can be calculated as
\begin{align}
E_{\textrm{HF}}(P_{\vec{k}})=\sum_{\vec{k}\in\textrm{mBZ}}\textrm{tr}\left[P_{\vec{k}}\left(H_{\textrm{cont}}(\vec{k})+H^T_{\textrm{HF}}\left(\vec{k},\frac{1}{2}P_{\vec{k}}\right)\right)\right]\label{HFener}.
\end{align}

\section{HF order parameter}
\label{HFparameter}
We first define the ingredients that we need to distinguish IVC, VP, and SP orders in Secs.~\ref{app:VP}-\ref{su2su2}, and then specify how we classify all states in Sec.~\ref{app:classHF}.

\subsection{Valley polarization}
\label{app:VP}
In this section, we detail our HF order parameter based on the HF correlator $P_{\vec{k}}$ defined in Appendix \ref{appenHFeq}. Before we begin, let us use a simplified notation: we denote the number of elements for a given set $Y$ as $N_{Y}$.

To define the HF order parameter for VP $\Delta_{\textrm{VP}}$ accounting for all possibilities among the three valley degrees of freedom, we first introduce the three matrices $\tau_{z_1}, \tau_{z_2}, \tau_{z_3}$ in valley space as
\begin{align}
\tau_{z_1} \equiv \left ( \begin{array}{cccc} 1 & 0 & 0\\
              0 & -1 &0 \\
             0 &0&0\\
             \end{array} \right), \ \ \tau_{z_2} \equiv \left ( \begin{array}{cccc} 0 & 0 & 0\\
              0 & 1 &0 \\
             0 &0&-1\\
             \end{array} \right), \ \ \tau_{z_3} \equiv \left ( \begin{array}{cccc} 1 & 0 & 0\\
              0 & 0 &0 \\
             0 &0&-1\\
             \end{array} \right) \nonumber.
\end{align}
We then define $\Delta_{\textrm{VP}}$ as
\begin{align}
\Delta_{\textrm{VP}-i}\equiv \frac{1}{N_{\textrm{mBZ}}}\sum_{\vec{k}\in\textrm{mBZ}}\sum_{s}\sum_{\eta}\langle c^{\dagger}_{\vec{k},s,\eta}[\tau_{z_i}]_{\eta,\eta'}c^{\pdagger}_{\vec{k},s,\eta}\rangle, \ \ \ \Delta_{\textrm{VP}} \equiv \underset{i\in \{1,2,3\}}{\textrm{Max}}|\Delta_{\textrm{VP}-i}|.
\end{align}\label{Eq.VPHF}

\subsection{Intervalley coherent order parameter}
In addition, to write the HF order parameter of IVC order $\Delta_{\textrm{IVC}}$, we first define $\tau_{x_1},\tau_{x_2},\tau_{x_3}, \tau_{x_4}$ in valley space and $s_t$ in spin space as

\begin{align}
\tau_{x_1} \equiv \left ( \begin{array}{cccc} 0 & 1 & 0\\
              1 & 0 &0 \\
             0 &0&0\\
             \end{array} \right), \ \ \tau_{x_2} \equiv \left ( \begin{array}{cccc} 0 & 0 & 0\\
              0 & 0 &1 \\
             0 &1&0\\
             \end{array} \right), \ \ \tau_{x_3} \equiv \left ( \begin{array}{cccc} 0 & 0 & 1\\
              0 & 0 &0 \\
             1 &0&0\\
             \end{array} \right), \ \ \tau_{x_4} \equiv \left ( \begin{array}{cccc} 0 & 0 & 1\\
              1 & 0 &0 \\
             0 &1&0\\
             \end{array} \right), \ \ s_{t} \equiv \left ( \begin{array}{cccc} 1& 1\\
              1 & 1\\
             \end{array} \right) \label{valleymat},
\end{align}
and define
\begin{align}
O_{\textrm{IVC}-j}\equiv \frac{1}{N_{\textrm{mBZ}}}\sum_{\vec{k}\in\textrm{mBZ}}\sqrt{\textrm{tr}[(P_{\vec{k}}\circ s_t\tau_{x_j})(P_{\vec{k}}\circ s_t\tau_{x_j})^{\dagger}]} \ \ (j=1,2,3) \label{Eq.IVC},
\end{align}
where we used Hadamard product $\circ$ for two matrices. Now, by defining $O_{\textrm{IVC}}$ as a descending ordered set of $\{O_{\textrm{IVC}-1}, O_{\textrm{IVC}-2}, O_{\textrm{IVC}-3} \}$, we introduce $\Delta_{\textrm{IVC}-i}$ as the $i$-th element of $O_{\textrm{IVC}}$, such that $\Delta_{\textrm{IVC}-1}\geq\Delta_{\textrm{IVC}-2}\geq\Delta_{\textrm{IVC}-3}$, and the corresponding IVC order parameter as $\Delta_{\textrm{IVC}}\equiv\Delta_{\textrm{IVC}-1}-\Delta_{\textrm{IVC}-3}$ which is zero for $\textrm{IVC}^3_\pm$ and non-zero for both NIVC orders.  

Lastly, we provide details of defining the IVC phase factor $\Delta_{\varphi}$. Here, we only consider the spin-diagonal term for simplicity. We first define the momentum set $\textrm{S}_s$ for each spin sector $s =\uparrow,\downarrow$ which contains all occupied momentum states that contribute to a finite IVC component as
\begin{align}
\textrm{S}_s\equiv\{\vec{k}\in\textrm{mBZ} \ |\ \sum_{\eta}\langle c^{\dagger}_{\vec{k},s,\eta}c_{\vec{k},s,\eta} \rangle>0, \ \  \sum_{\eta}\langle c^{\dagger}_{\vec{k},s,\eta}c_{\vec{k},s,\eta+1} \rangle  \neq 0\}\label{Eq.momset}.
\end{align}
Using \equref{Eq.momset}, we define $\Delta_{\varphi}(\vec{k})$ and $\Delta_{\varphi}$ as
\begin{subequations}
    \begin{align}
    \begin{split}\label{eq.Deltaphi1}
N_{\textrm{S}_{\uparrow}}=  N_{\textrm{S}_{\downarrow}}=0 \rightarrow \Delta_{\varphi}(\vec{k})=0, \ \  \Delta_{\varphi}= 0 
    \end{split} \\
    \begin{split}
     N_{\textrm{S}_{\uparrow}}>0, N_{\textrm{S}_{\downarrow}}=0 \rightarrow \Delta_{\varphi}(\vec{k})\equiv \sum_{\eta,\eta'}\textrm{Arg}[\langle c^{\dagger}_{\vec{k},\uparrow,\eta}[\tau_{x_4}]_{\eta,\eta'}c_{\vec{k},\uparrow,\eta'}\rangle], \ \  \Delta_{\varphi}= \frac{1}{N_{\textrm{S}_{\uparrow}}}\sum_{\vec{k}\in \text{S}_{\uparrow}}\Delta_{\varphi}(\vec{k})
    \end{split}\\
    \begin{split}
     N_{\textrm{S}_{\uparrow}}=0, N_{\textrm{S}_{\downarrow}}>0 \rightarrow \Delta_{\varphi}(\vec{k})\equiv \sum_{\eta,\eta'}\textrm{Arg}[\langle c^{\dagger}_{\vec{k},\downarrow,\eta}[\tau_{x_4}]_{\eta,\eta'}c_{\vec{k},\downarrow,\eta'}\rangle], \ \  \Delta_{\varphi}= \frac{1}{N_{\textrm{S}_{\downarrow}}}\sum_{\vec{k}\in \text{S}_{\downarrow}}\Delta_{\varphi}(\vec{k})
    \end{split}\\
    \begin{split}\label{eq.Deltaphi4}
    N_{\textrm{S}_{\uparrow}}>0, N_{\textrm{S}_{\downarrow}}>0 \rightarrow \Delta_{\varphi}(\vec{k})\equiv \sum_s\sum_{\eta,\eta'}\textrm{Arg}[\langle c^{\dagger}_{\vec{k},s,\eta}[\tau_{x_4}]_{\eta,\eta'}c_{\vec{k},s,\eta'}\rangle], \ \  \Delta_{\varphi}= \sum_{\vec{k}\in \text{S}_{s}}\sum_s\frac{1}{2N_{\textrm{S}_{s}}}\Delta_{\varphi}(\vec{k}),
    \end{split}
    \end{align}\label{IVCphase}\end{subequations}
with $\Delta_{\varphi}(\vec{k})$ computed by modulus $2\pi$, where $0\leq\Delta_{\varphi}(\vec{k})<2\pi$. Note that $\Delta_{\varphi}$ in our HF calculation corresponds to $\varphi_{123}$ defined in \equref{DescriptionOfStates}.

\subsection{Spin polarization} \label{su2su2}
Due to the presence of $[\textrm{SU(2)}]_\text{S}^3$ spin rotation symmetry (acting on each valley degree of freedom), states with different spin orientations can be transformed into each other. Thus, to define an SP order parameter, we would like to distinguish between states that are equivalent to a spin-unpolarized state and those which cannot become unpolarized (by means of a $[\textrm{SU(2)}]_\text{S}^3$ transformation). To 
this end, we determine the state $P_{\vec{k}}$ with the maximal and minimal spin polarization among the class of states connected via $[\textrm{SU(2)}]_\text{S}^3$ transformations. The SU(2)$_\text{S}$ spin transformation matrix $R_{\textrm{SU(2)}_\text{S}}$ for each valley degree of freedom $ \eta$ can be parametrized as $R_{\textrm{SU(2)}_\text{S},\eta}=e^{i\psi_{1,\eta}s_z}e^{i\psi_{2,\eta}s_x}e^{i\psi_{3,\eta}s_z}$ with three Euler angles $(\psi_{1,\eta},\psi_{2,\eta},\psi_{3,\eta})$. By transforming $P_{\vec{k}}\rightarrow P'_{\vec{k}}\equiv\langle c'^{\dagger}_{\vec{k}}c'_{\vec{k}} \rangle$ via $P'_{\vec{k}}=R^{\dagger}_{\textrm{SU(2)}_\text{S},1}R^{\dagger}_{\textrm{SU(2)}_\text{S},2}R^{\dagger}_{\textrm{SU(2)}_\text{S},3}P_{\vec{k}}R_{\textrm{SU(2)}_\text{S},3}R_{\textrm{SU(2)}_\text{S},2}R_{\textrm{SU(2)}_\text{S},1}$, we define the two spin-polarization order parameters $\Delta_{\textrm{SP}_1}$ and $\Delta_{\textrm{SP}_2}$ as
\begin{subequations}
\begin{align}
\begin{split}
\Delta_{\textrm{SP}_1}\equiv\frac{1}{N_{\textrm{mBZ}}}\underset{\{\psi_{j,\eta}\}}{\textrm{Min}}\sum_{\alpha=x,y,z}\left|\sum_{\vec{k\in \textrm{mBZ}}}\sum_{s}\sum_{\eta}\langle c'^{\dagger}_{\vec{k},s,\eta}[s_{\alpha}]_{s,s'}c'_{\vec{k},s',\eta}\rangle \right|, \end{split}\\
\begin{split}
\Delta_{\textrm{SP}_2}\equiv\frac{1}{N_{\textrm{mBZ}}}\underset{\{\psi_{j,\eta}\}}{\textrm{Max}}\sum_{\alpha=x,y,z}\left|\sum_{\vec{k\in \textrm{mBZ}}}\sum_{s}\sum_{\eta}\langle c'^{\dagger}_{\vec{k},s,\eta}[s_{\alpha}]_{s,s'}c'_{\vec{k},s',\eta}\rangle \right|.
\end{split}
\end{align}\label{SPmaxmin}
\end{subequations}

\subsection{Classification of the HF ground states}
\label{app:classHF}
With these definitions, we classify the possible orders with the following criteria 
\begin{subequations}
    \begin{align}
    \begin{split}
\textrm{VP}_{\uparrow}: \Delta_{\textrm{SP}_1}=1, \  \Delta_{\textrm{VP}}>0, \  \Delta_{\textrm{IVC}-i}=0\nonumber
    \end{split} \\
      \begin{split}
\textrm{VP}_{\uparrow \downarrow}: \Delta_{\textrm{SP}_1}=0, \  \Delta_{\textrm{VP}}>0, \  \Delta_{\textrm{IVC}-i}=0\nonumber
    \end{split} \\
    \begin{split}
\textrm{SP}: \Delta_{\textrm{SP}_1}>0, \  \Delta_{\textrm{VP}}=0, \  \Delta_{\textrm{IVC}-i}=0\nonumber
    \end{split} \\
    \begin{split}
     \textrm{Sym}: \Delta_{\textrm{SP}_2}=0, \ \Delta_{\textrm{VP}}=0,\  \Delta_{\textrm{IVC}-i}=0\nonumber
    \end{split}\\
    \begin{split}    \textrm{IVC}^3_+:\Delta_{\textrm{VP}}=0, \  \Delta_{\textrm{IVC}-i}>0, \  \Delta_{\textrm{IVC}}=0, \ \Delta_{\varphi}=0 \nonumber 
    \end{split} \\
     \begin{split}
   \textrm{IVC}^3_-: \Delta_{\textrm{VP}}=0, \  \Delta_{\textrm{IVC}-i}>0, \  \Delta_{\textrm{IVC}}=0, \ \Delta_{\varphi}=\pi \nonumber
    \end{split} \\
    \begin{split}
      \textrm{NIVC}^1: \Delta_{\textrm{VP}}>0, \  \Delta_{\textrm{IVC}-1}>0, \ \Delta_{\textrm{IVC}-2}= \Delta_{\textrm{IVC}-3}=0 \nonumber
    \end{split} \\
    \begin{split}
      \textrm{NIVC}^3: \Delta_{\textrm{VP}}>0, \ \Delta_{\textrm{IVC}\pm1,2}^2+\Delta_{\textrm{IVC}\pm2,3}^2>0, \  \Delta_{\varphi}(\vec{k})=0,\pi \nonumber
    \end{split}
  \end{align}
\end{subequations}
where we additionally defined $\Delta_{\textrm{IVC}\pm i,j}\equiv\Delta_{\textrm{IVC}-i}\pm\Delta_{\textrm{IVC}-j}$. We note that in our numerics, we also see that the mixture of NIVC$^3$ between $\Delta_{\varphi}(\vec{k})=0$ and $\Delta_{\varphi}(\vec{k})=\pi$ is a ground state, which corresponds to a non-quantized value of $\Delta_{\varphi}$. For the IVC$^3_+$ (IVC$^3_-$) state, $\Delta_{\varphi} = 0 \  (\Delta_{\varphi}=\pi)$ directly imposes $\Delta_{\varphi}(\vec{k})=0 \ (\Delta_{\varphi} (\vec{k})=\pi)$ for all $\vec{k}\in\text{mBZ}$.

\section{Additional HF data}\label{HFextradata}
\begin{figure*}[t]
	\centering
	\includegraphics[width= \linewidth]{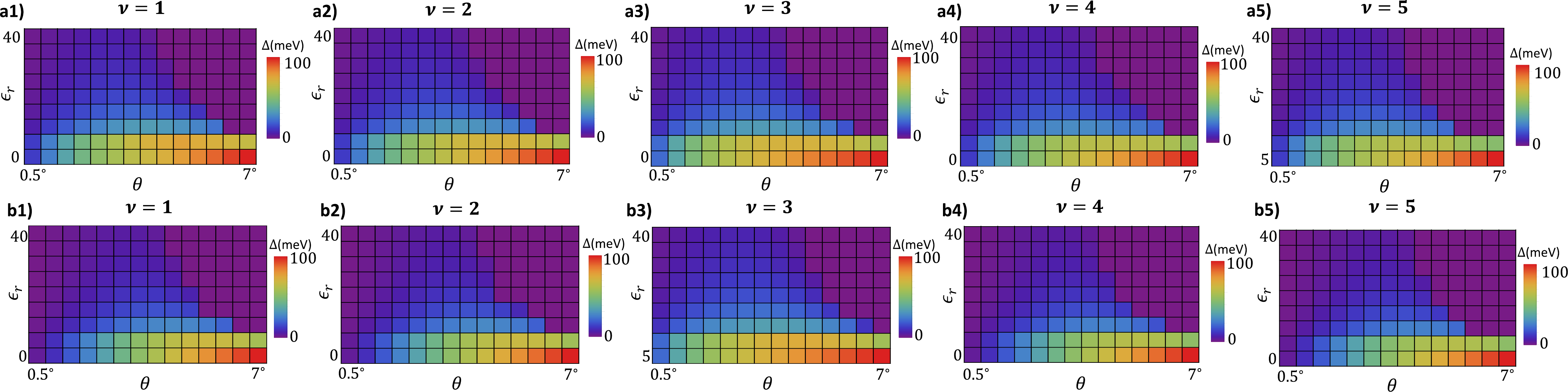}
	\caption{{\bf HF band-gap of AA and AB stacking}
\justifying
   Indirect band-gap $\Delta$ obtained from HF as function of twist angle $\theta$ and relative permittivity $\epsilon_r$ for \textbf{(a1 - a5)} AA stacking and \textbf{(b1 - b5)} AB stacking at each integer filling. The system size is $12\times12$.}
	\label{Fig4:HFbandgap}
\end{figure*}

\begin{figure*}[t]
	\centering
	\includegraphics[width= \linewidth]{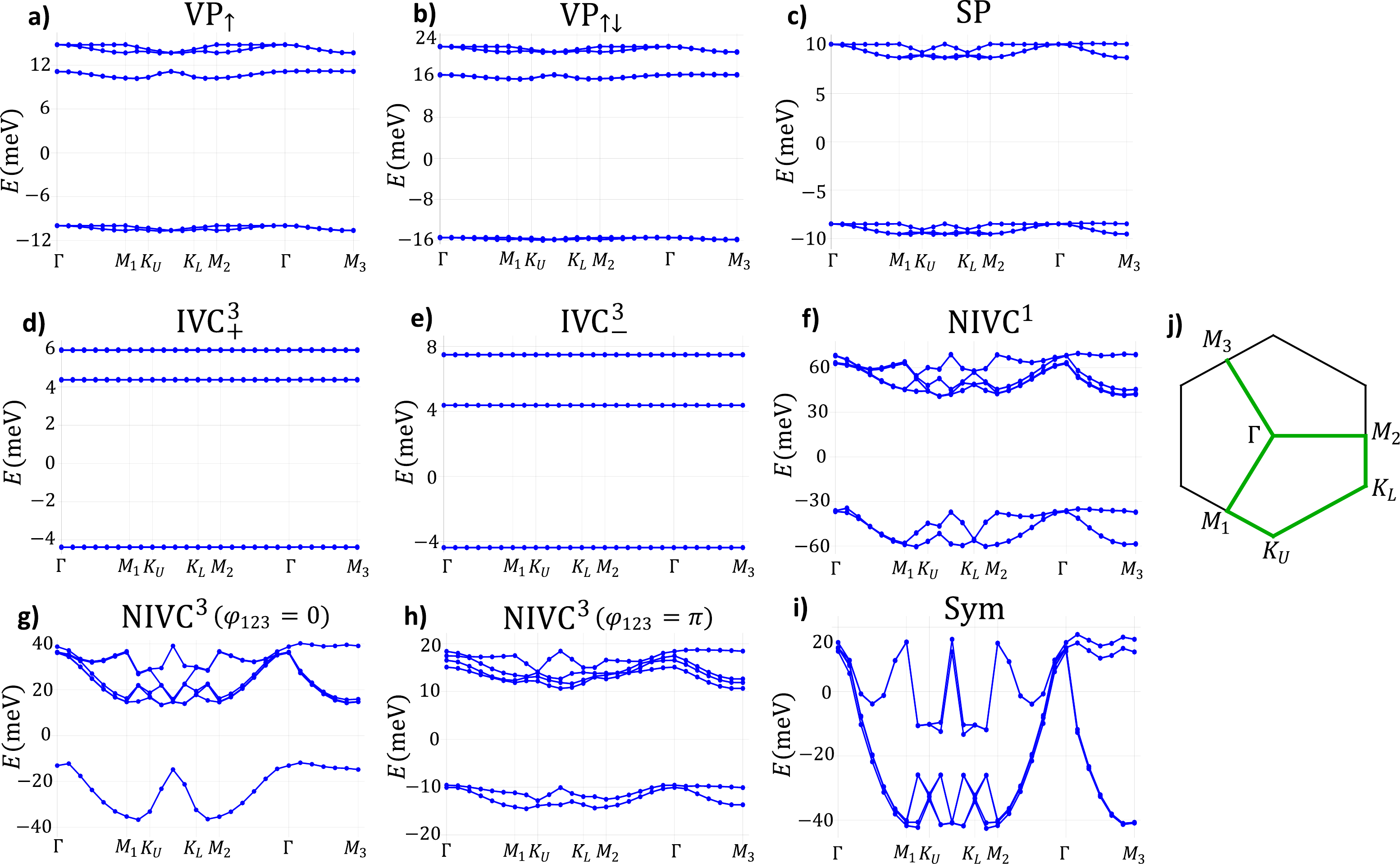}
	\caption{{\bf HF band structure of AA stacking for each ground states}
\justifying
  HF band structure in the case of AA stacking along the high symmetry momentum path ($\Gamma-M_1-K_U-K_L-M_2-\Gamma-M_3$) described in (\textbf{j}) for each ground state: (\textbf{a})  $\textrm{VP}_{\uparrow} \ (\nu=2, \theta=3^{\circ}, \epsilon_r=15)$, (\textbf{b})  $\textrm{VP}_{\uparrow\downarrow} \ (\nu=2, \theta=3^{\circ}, \epsilon_r=15)$, (\textbf{c})  $\textrm{SP} \ (\nu=3, \theta=3^{\circ}, \epsilon_r=20)$, (\textbf{d})  $\textrm{IVC}^3_+ \ (\nu=1, \theta=1^{\circ}, \epsilon_r=15)$, (\textbf{e})  $\textrm{IVC}^3_- \ (\nu=2, \theta=1^{\circ}, \epsilon_r=15)$, (\textbf{f})  $\textrm{NIVC}^1 \ (\nu=2, \theta=6^{\circ}, \epsilon_r=5)$, (\textbf{g}) $\textrm{NIVC}^3$ $(\varphi_{123}=0,\nu=1, \theta=6^{\circ}, \epsilon_r=10)$, (\textbf{h}) $\textrm{NIVC}^3$ $(\varphi_{123}=\pi, \nu=2, \theta=4^{\circ}, \epsilon_r=15)$ and (\textbf{i}) $\textrm{Sym}$ $(\nu=5, \theta=7^{\circ}, \epsilon_r=20)$. The Fermi level is located at zero energy in every figure. The system size is $12\times12$. 
	}
	\label{Fig5:HFbandstructure}
\end{figure*}

\begin{figure*}[t]
	\centering
	\includegraphics[width= \linewidth]{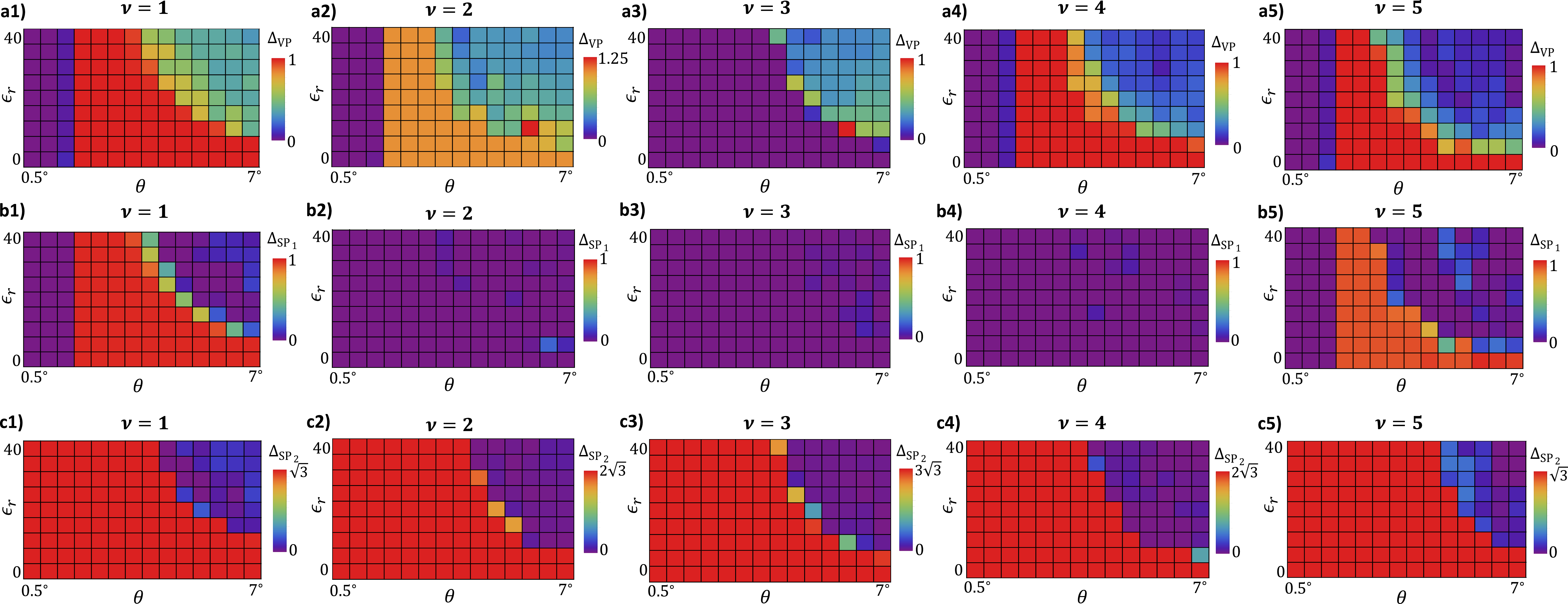}
	\caption{{\bf HF order parameter for valley polarization and spin-polarization in the case of AA stacking.}
\justifying
  Dependence on twist angle $\theta$ and relative permittivity $\epsilon_r$ of \textbf{(a1-a5)} valley polarization $\Delta_{\textrm{VP}}$,  
	\textbf{(b1-b5)} spin polarization $\Delta_{\textrm{SP}_1}$, and \textbf{(c1-c5)} $\Delta_{\textrm{SP}_2}$ at each integer filling. The order parameters are defined in \equref{Eq.VPHF} and \ref{SPmaxmin}. The system size is $12\times12$.} 
	\label{Fig5:HFVPorder}
\end{figure*}

\begin{figure*}[t]
	\centering
	\includegraphics[width= \linewidth]{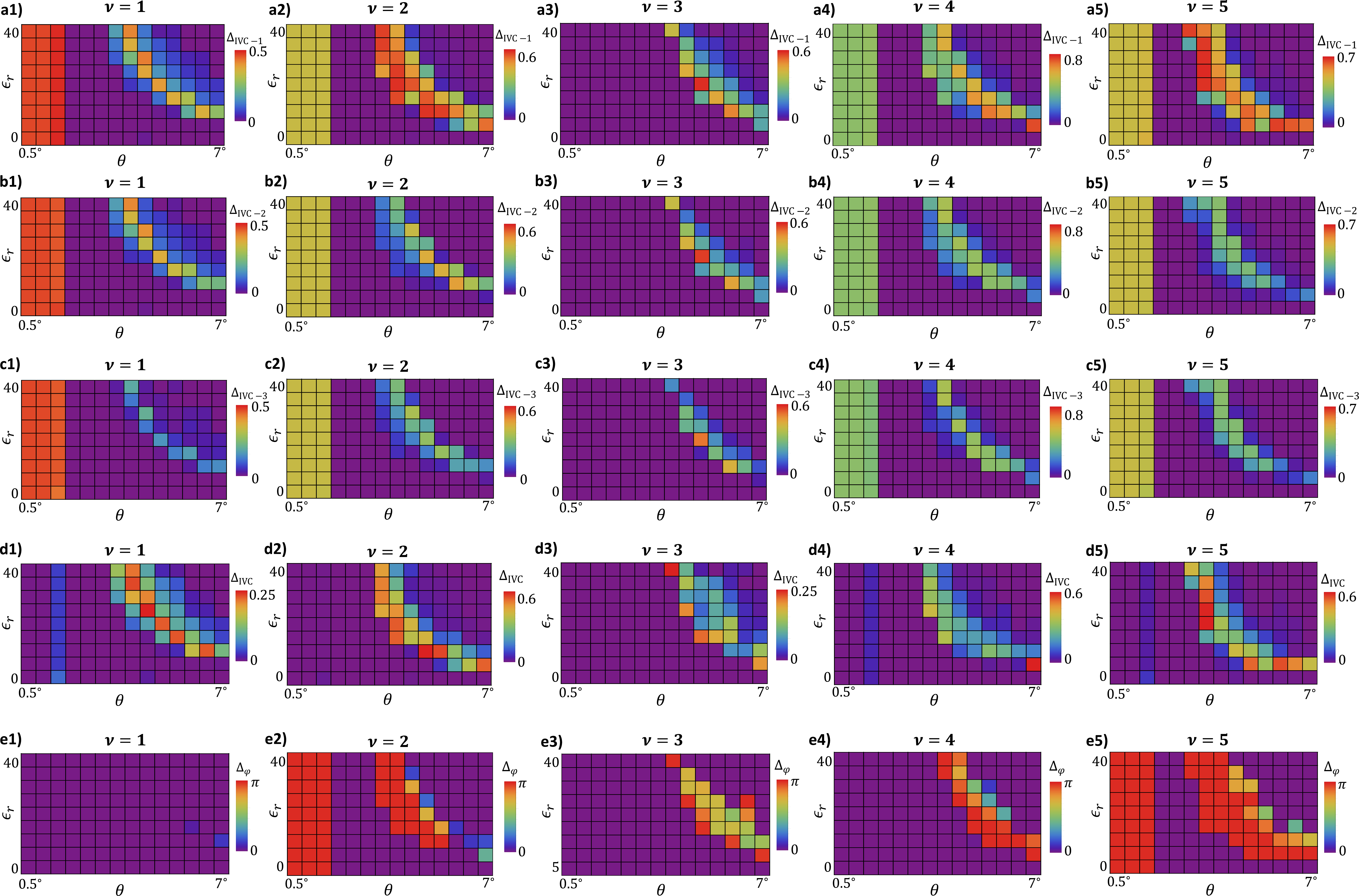}
	\caption{{\bf HF order parameter for intervalley coherent order in the case of AA stacking.}
\justifying
  Dependence on twist angle $\theta$ and relative permittivity $\epsilon_r$ dependence of (\textbf{a1-a5}) $\Delta_{\textrm{IVC}-1}$, (\textbf{b1-b5}) $\Delta_{\textrm{IVC}-2}$, (\textbf{c1-c5}) $\Delta_{\textrm{IVC}-3}$, (\textbf{d1-d5}) $\Delta_{\textrm{IVC}}$, and (\textbf{e1-e5}) $\Delta_{\varphi}$ at each integer filling. The order parameters are defined below Eq.~\eqref{Eq.IVC} and in Eqs. \eqref{eq.Deltaphi1}-\eqref{eq.Deltaphi4}. The system size is $12\times12$.
	}
	\label{Fig10:HFIVCorder}
\end{figure*}

\begin{figure*}[t]
	\centering
	\includegraphics[width= \linewidth]{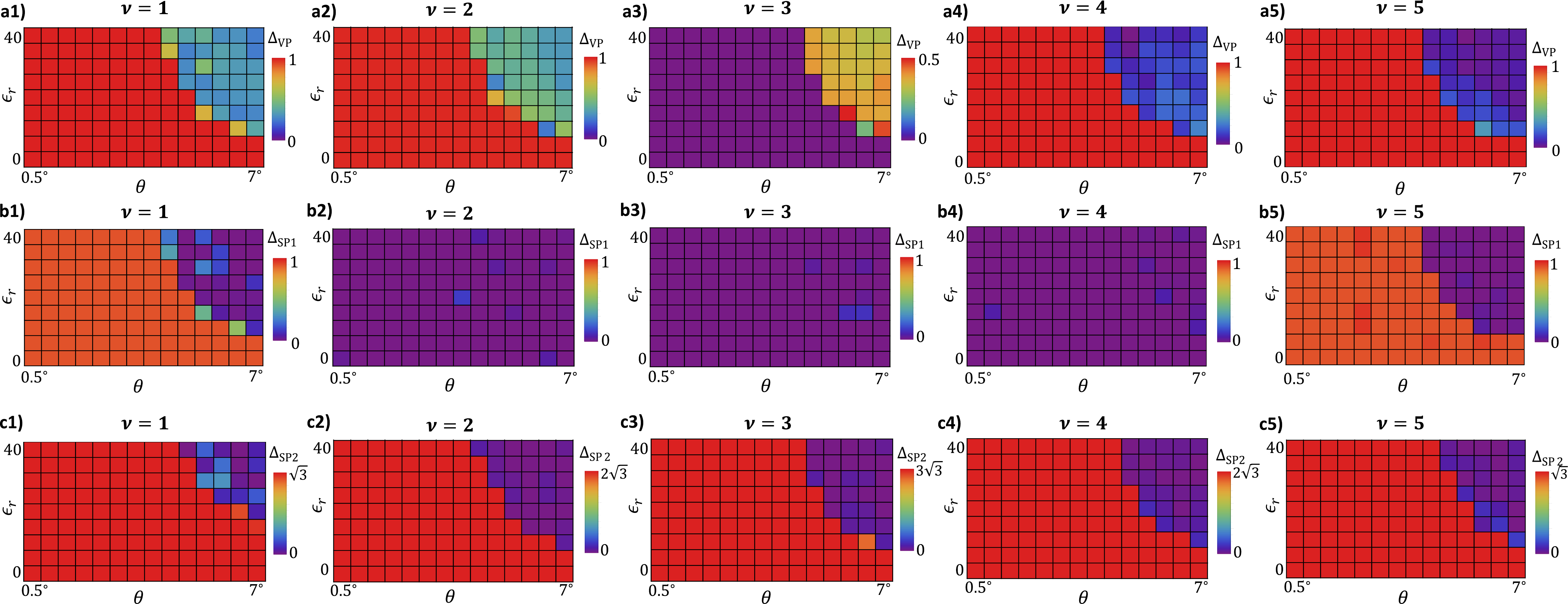}
	\caption{{\bf HF order parameters for valley polarization and spin polarization in the case of AB stacking.}
\justifying
  Dependence on twist angle $\theta$ and relative permittivity $\epsilon_r$ of \textbf{(a1-a5)} valley polarization $\Delta_{\textrm{VP}}$ 
	\textbf{(b1-b5)} spin polarization$\Delta_{\textrm{SP}_1}$, and \textbf{(c1-c5)} $\Delta_{\textrm{SP}_2}$ at each integer filling. The order parameters are defined in \equref{Eq.VPHF} and \ref{SPmaxmin}. The system size is $12\times12$.} 
	\label{Fig10:ABHFVPorder}
\end{figure*}

\begin{figure*}[t]
	\centering
	\includegraphics[width= \linewidth]{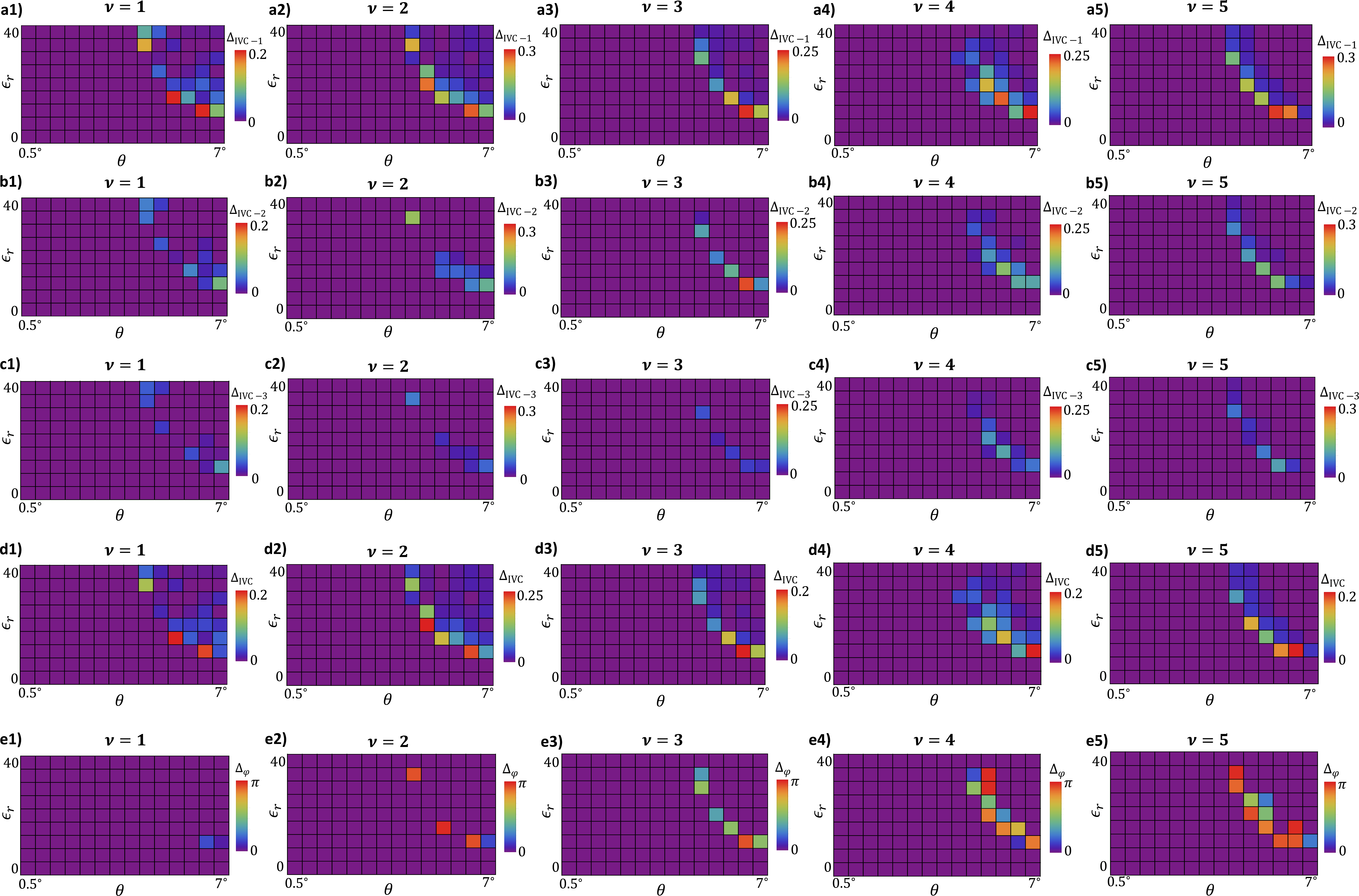}
	\caption{{\bf HF order parameters for intervalley coherent order in the case of AB stacking.}
\justifying
   Dependence on twist angle $\theta$ and relative permittivity $\epsilon_r$ of (\textbf{a1-a5}) $\Delta_{\textrm{IVC}-1}$, (\textbf{b1-b5}) $\Delta_{\textrm{IVC}-2}$, (\textbf{c1-c5}) $\Delta_{\textrm{IVC}-3}$, (\textbf{d1-d5}) $\Delta_{\textrm{IVC}}$, and (\textbf{e1-e5}) $\Delta_{\varphi}$ for each integer filling.  The order parameters are defined below Eq.~\eqref{Eq.IVC} and in Eqs. \eqref{eq.Deltaphi1}-\eqref{eq.Deltaphi4}. The system size is $12\times12$.} 
	\label{Fig10:ABHFIVCorder}
\end{figure*}

In this section, we show additional HF data of the three-valley moir\'e system. Figure \ref{Fig4:HFbandgap} shows the dependence on twist angle $\theta$ and on the relative permittivity $\epsilon_r$ of the HF indirect band gap $\Delta$ between the highest occupied band and the lowest unoccupied band for AA and AB stacking at each integer filling. As we mentioned in the main text, we find a metal-to-insulator transition at larger values of $\theta$ and $\epsilon_r$, where the indirect band gap closes. In Fig.~\ref{Fig5:HFbandstructure} we show examples of the HF band structure for the symmetric and each ordered HF state in the case of AA stacking along the depicted high symmetry momentum path. For the chosen angle of $\theta=3^\circ$, the ordered states clearly lead to an insulating band structure. In the symmetric case, there is a strong band renormalization which increases the bandwidth of the lowest bands. In Figs.~\ref{Fig5:HFVPorder}-\ref{Fig10:ABHFIVCorder}, we show the HF order parameters defined in Eqs.~(\ref{Eq.VPHF})-(\ref{SPmaxmin}) as function of twist angle $\theta$ and the relative permittivity $\epsilon_r$ at integer fillings for AA and AB stacking, respectively. Figs.~\ref{Fig5:HFVPorder} and \ref{Fig10:ABHFVPorder} contain valley polarization $\Delta_{\textrm{VP}}$ and spin polarization $\Delta_{\textrm{SP}_1}, \Delta_{\textrm{SP}_2}$. Figures~\ref{Fig10:HFIVCorder} and \ref{Fig10:ABHFIVCorder} contain $\Delta_{\textrm{IVC}-i} \ (i=1,2,3)$, $\Delta_{\textrm{IVC}}$ and $\Delta_{\varphi}$. These data are used to construct the phase diagram in the main text.

\section{Checking the FMC and U(6) limit}\label{App:FMCSU3}
To check the FMC of both AA and AB stacking, where we impose the form factor $\Lambda(\vec{k}, \vec{G})$ to be independent of $\vec{k}\in\textrm{mBZ}$ and the valley degree of freedom $\eta$ for a given moir\'e reciprocal vector $\vec{G}$, such that $\Lambda_{\eta}(\vec{k}, \vec{G})=f(\vec{G})$, we numerically compute the form factor width $X$, which is defined as 
 \begin{align}
 \begin{split}
 X \equiv  \frac{1}{108N^2}\sum_{\eta}\sum_{\vec{k,k}'\in\textrm{mBZ}}\sum_{\vec{G}\in\textrm{RL}}\left(|\Lambda_{\eta}(\vec{k},\vec{G})-    \Lambda_{\eta+1}(\vec{k}',\vec{G})|+|\Lambda_{\eta}(\vec{k},\vec{G})-    \Lambda_{\eta}(\vec{k}',\vec{G})|\right).
 \end{split}\label{flatmetric} 
\end{align}
We use a cutoff by only including reciprocal lattice vectors with  $0<|\vec{G}|\leq2\sqrt{3}k_{\theta}$ in the summation and denote the system size by $N$. Note that we consider the normalization factor $1/108$ from the number of reciprocal vectors $(=18)$, three pairs of equal $(=3)$ and unequal valleys $(=3)$, which leads to $X\leq 2$. We also define $(\eta=4)\equiv(\eta=1)$ for the valley index.

\begin{figure*}[t]
	\centering
	\includegraphics[width= 0.5\columnwidth, height=0.28\columnwidth]{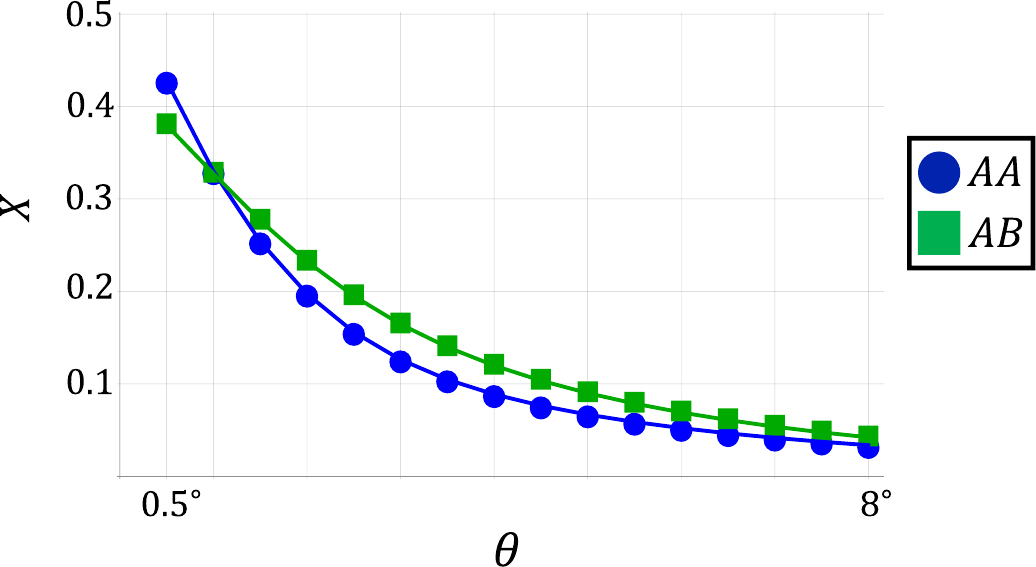}
	\caption{{\bf FMC violation of form factor.}
\justifying
Form factor width $X$ defined in \equref{flatmetric} as a function of twist angle $\theta$ for both AA and AB stacking.}
	\label{Fig12:flatbandmetric}
\end{figure*}

In addition, to investigate the role of U(6) symmetry-breaking in the violation of the FMC, we define $Y$ 
\begin{align}
  \begin{split}
 Y \equiv  \frac{1}{108N}\sum_{\eta}\sum_{\vec{k}\in\textrm{mBZ}}\sum_{\vec{G}\in \text{RL}}(|\Lambda_{\eta}(\vec{k},\vec{G})-\Lambda_{\eta+1}(\vec{k},\vec{G})|), \ 
 \end{split} 
 \end{align}  
and compare $X$ and $Y$. 
Figure \ref{Fig12:flatbandmetric} shows the twist angle dependence of $X$ for both AA and AB stacking. We clearly see that there is a large violation of the FMC  for both AA and AB stacking when the twist angle is small. We also observe that AA stacking leads to a slightly larger violation of the FMC than AB stacking for the smallest twist angle we investigate. We find numerically that $X$ and $Y$ coincide, indicating that the violation of the FMC comes from the valley dependence, which also means that U(6) is simultaneously broken. 
We point out that for $\theta=1.5^{\circ}$, even though AA stacking leads to a smaller $X$ than AB stacking, we see that the $\text{IVC}^3_{\pm}$ state becomes the ground state for AA stacking, whereas VP is the ground state for AB stacking.  
We note in this context that the FMC is a functional equation and we use the simple scalar quantities $X,Y$ as characteristics to investigate trends, which, however cannot reveal the full functional form. Overall, our strong coupling analysis and these numerical results for the FMC characteristics agree qualitatively well and provide an explanation why we see the $\text{IVC}^3_{\pm}$ state as the ground state at small twist angles only for AA stacking. Moreover, as we have discussed above in \appref{MomentumDependentIVC}, the momentum dependence beyond the perturbative regime might also play a role for the energetics of the $\text{IVC}^3_{\pm}$ state for AA stacking.

\section{Effect of the intravalley $C_{2z}$ symmetry}\label{effectintra}
\begin{figure*}[t]
	\centering
	\includegraphics[width= \linewidth]{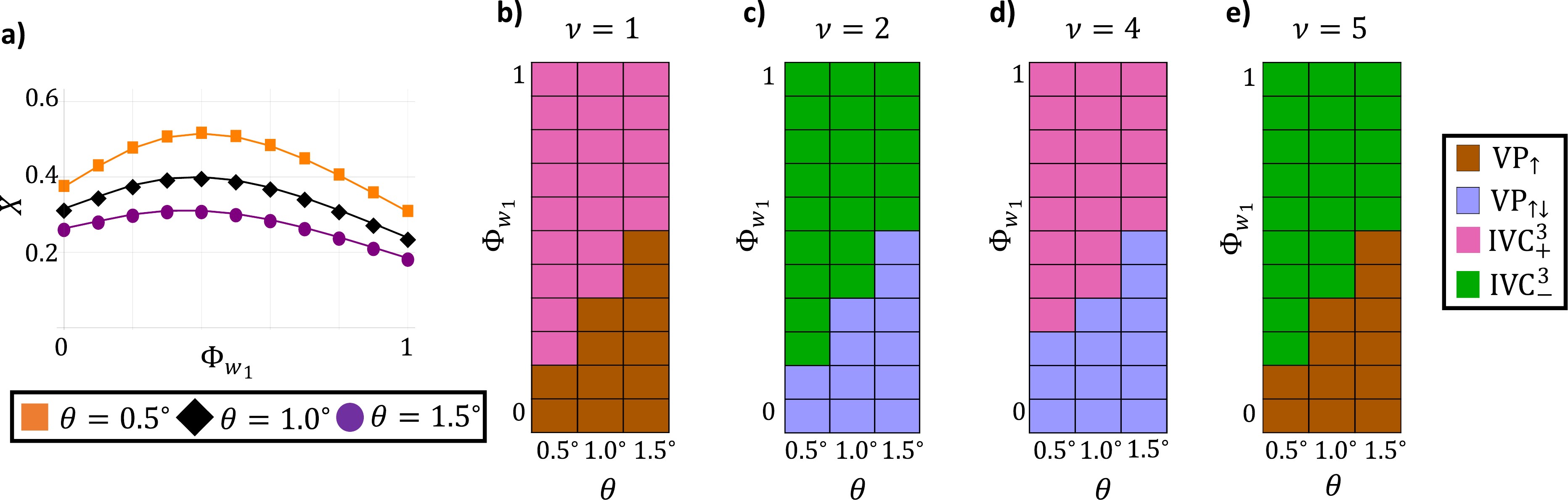}
	\caption{{\bf Effect of breaking intravalley $C_{2z}$ symmetry.}
\justifying
  (\textbf{a}) Form factor width $X$ defined in \equref{flatmetric} as a function of $C_{2z}$ symmetry-breaking parameter $\Phi_{w_1}=\textrm{Im}[w_1]/\textrm{Re}[w_1]$ for twist angle $\theta=0.5^{\circ}, 1.0^{\circ}, 1.5^{\circ}$. (\textbf{b-e}) HF phase diagrams as function of $\Phi_{w_1}$ for integer fillings $\nu=1,2,4,5$, with relative permittivity $\epsilon_r=5$. Here, we fixed $\textrm{Re}[w_1]=88.0 \ \textrm{meV},w_2 = -18.94 \ \textrm{meV}$. The system size is $12\times12$.}
	\label{Fig11:C2zbreak}
\end{figure*}

In this section, we focus on how the intravalley $C_{2z}$ symmetry affects the energetic competition between $\textrm{IVC}^{3}_{\pm}$ and $\{\textrm{VP}_{\uparrow}, \textrm{VP}_{\uparrow\downarrow}\}$ for small twist angles $0.5^{\circ}\leq\theta\leq1.5^{\circ}$. The presence of $C_{2z}\Theta$ symmetry constrains the interlayer tunneling $w_1$ to be real valued. To systematically break the intravalley $C_{2z}$ symmetry, we first fix $\textrm{Re}[w_1]=88.80 \ \textrm{meV}, w_2=-18.94 \ \textrm{meV}$, and then tune the following ratio $0\leq\Phi_{w_1}\equiv \textrm{Im}[w_1]/\textrm{Re}[w_1]\leq1$. Note that the AA stacking we considered for the HF calculation in the previous sections corresponds to $\Phi_{w_1}=66.38/88.8\approx0.75$. Figure~\ref{Fig11:C2zbreak}(a) shows the form factor width $X$ with respect to $\Phi_{w_1}$ for different twist angles $\theta$. In addition, Figs.\ \ref{Fig11:C2zbreak}(b-e) show the HF phase diagrams with respect to the twist angle $\theta$ and $\Phi_{w_1}$ for each integer filling $\nu = 1,2,4,5$ (we do not consider $\nu=3$, since we see a SP state as the HF ground state regardless of the stacking structure). We fix the relative permittivity to $\epsilon_r=5$ in these plots. As expected, we see the $\{\textrm{VP}_{\uparrow}, \textrm{VP}_{\uparrow\downarrow}\} \rightarrow \textrm{IVC}^3_{\pm}$ the phase transition with respect to $\Phi_{w_1}$. The transition occurs for smaller $\Phi_{w_1}$ the smaller $\theta$. 
In addition, we see a non-monotonous behavior of $X$ as a function of  $\Phi_{w_1}$; it increases up to $\Phi_{w_1}\approx 0.4$ and then decreases. In conjunction with our earlier conclusion (see App.~\ref{App:FMCSU3}), this indicates that it is the combination of FMC violation and momentum dependence of the IVC order parameter that allows the $\textrm{IVC}_\pm^3$ state to be energetically favored against VP.

\end{appendix}

\end{document}